\documentclass[11pt]{article}

\usepackage{amsfonts}
\usepackage{amsmath}
\usepackage{amssymb}
\usepackage{mathrsfs}
\usepackage{amsthm}
\usepackage[english]{babel}
\usepackage[T1]{fontenc}
\usepackage[latin1]{inputenc}
\usepackage{epsfig}
\usepackage{epstopdf}
\usepackage{textcomp}
\usepackage{graphicx}
\usepackage{booktabs}

\numberwithin{equation}{section}
\newcommand{\clearemptydoublepage}{\newpage{\pagestyle{empty}\cleardoublepage}}

\def\a{\alpha}
\def\b{\beta}
\def\g{\gamma}

\def\d{\delta}

\def\ep{\epsilon}

\def\z{\zeta}
\def\e{\eta}

\def\la{\lambda}
\def\La{\Lambda}
\def\m{\mu}
\def\n{\nu}

\def\vp{\varpi}
\def\r{\rho}

\def\s{\sigma}

\def\t{\tau}

\def\vf{\varphi}
\def\ch{\chi}

\def\o{\omega}

\def\tt{\tilde{\tau}}

\def\tvp{\tilde{\varpi}}

\newcommand{\ti}[1]{\tilde{#1}}

\newcommand{\Xd}{X^{\cdot}}
\newcommand{\xid}{\xi^{\cdot}}
\newcommand{\chd}{\chi^{\cdot}}

\newcommand{\xd}{x^{\cdot}}
\newcommand{\vfd}{\varphi^{\cdot}}

\newcommand{\bn}{\bar{n}}

\newcommand{\bvf}{\bar{\varphi}}
\newcommand{\bvfd}{\bar{\varphi}^{\cdot}}

\newcommand{\bpi}{\bar{\pi}}

\newcommand{\hxi}{\hat{\xi}}
\newcommand{\hxid}{\hat{\xi}^{\cdot}}

\newcommand{\h}[1]{\hat{#1}}
\newcommand{\hp}{\hat{\pi}}

\providecommand{\abs}[1]{\lvert#1\rvert}

\newcommand{\de}{\partial}
\newcommand{\half}{\frac{1}{2}}
\newcommand{\third}{\frac{1}{3}}

\newcommand{\p}{\prime}

\newcommand{\beq}{\begin{equation}}
\newcommand{\eeq}{\end{equation}}
\newcommand{\beqnn}{\begin{equation*}}
\newcommand{\eeqnn}{\end{equation*}}
\newcommand{\bea}{\begin{eqnarray}}
\newcommand{\eea}{\end{eqnarray}}
\newcommand{\nn}{\nonumber}
\newcommand{\noi}{\noindent}

\newcommand{\bq}{\beta^{2}}

\newcommand{\xim}{\xi^{\mu}}
\newcommand{\chm}{\chi^{\mu}}

\newcommand{\mbfG}{\mathbf{G}}

\newcommand{\tmbfg}{\tilde{\mathbf{g}}}
\newcommand{\tmbfK}{\tilde{\mathbf{K}}}
\newcommand{\tmbfG}{\tilde{\mathbf{G}}}
\newcommand{\tmbfT}{\tilde{\mathbf{T}}}
\newcommand{\mcal}[1]{\mathcal{#1}}
\newcommand{\mcalT}{\mathcal{T}}
\newcommand{\mcalC}{\mathcal{C}}

\newcommand{\mcalL}{\mathcal{L}}

\newcommand{\mscr}[1]{\mathscr{#1}}
\newcommand{\mscrT}{\mathscr{T}}

\newcommand{\mscrI}{\mathscr{I}}

\newcommand{\ded}[1]{\de_{#1}}

\newcommand{\dem}{\de_{\m}}

\newcommand{\demden}{\de_{\m} \de_{\n}}
\newcommand{\boxf}{\Box_{4}}

\newcommand{\boxfi}{\Box_{5}}
\newcommand{\boxs}{\Box_{6}}

\newcommand{\dexim}{\de_{\xi^{\m}}}

\newcommand{\deximxin}{\de_{\xi^{\m}} \de_{\xi^{\n}}}

\newcommand{\dechmchn}{\de_{\chi^{\m}} \de_{\chi^{\n}}}

\newcommand{\evb}{\Big\rvert_{X^{\cdot} = \bar{\varphi}^{\cdot}(\xi^{\cdot})}}

\newcommand{\evbh}{\Big\rvert_{X^{\cdot} = \bar{\varphi}^{\cdot}(\hat{\xi}^{\cdot})}}
\newcommand{\evbhsi}{\Big\rvert_{\bar{\varphi}^{\cdot}(\hat{\xi}^{\cdot})}}

\newcommand{\evbbhsi}{\bigg\rvert_{\bar{\varphi}^{\cdot}(\hat{\xi}^{\cdot})}}

\newcommand{\Msf}{M_{6}^{4}}
\newcommand{\Mft}{M_{5}^{3}}
\newcommand{\Mfs}{M_{4}^{2}}

\newcommand{\Zp}{Z^{\prime}}

\newcommand{\Yp}{Y^{\prime}}

\newcommand{\mZ}{\mathscr{Z}}
\newcommand{\mY}{\mathscr{Y}}

\newcommand{\mcalZpq}{{\mathcal{Z}^{\prime}}^2}

\newcommand{\mcalYpq}{{\mathcal{Y}^{\prime}}^2}

\newcommand{\bla}{\bar{\la}}
\newcommand{\blac}{\bar{\la}_{c}}
\newcommand{\dla}{\d \! \la}

\newcommand{\dvf}{\delta \! \varphi}
\newcommand{\hdvf}{\delta \! \hat{\varphi}}

\newcommand{\dvfn}{\delta \! \varphi_{\!_{\perp}}}
\newcommand{\hdvfn}{\delta \! \hat{\varphi}_{\!_{\perp}}}

\newcommand{\dvfp}{\delta \! \varphi_{\shortparallel}}

\newcommand{\bv}{\bar{v}}
\newcommand{\dv}{\delta v}

\newcommand{\bep}{\bar{\epsilon}}
\newcommand{\bN}{\bar{N}}
\newcommand{\bNq}{\bar{N}^{2}}

\newcommand{\bNy}{\bar{N}_{y}}

\newcommand{\bNyp}{\bar{N}_{y}^{\p}}
\newcommand{\debz}{\partial_{\bar{z}}}
\newcommand{\debzq}{\partial^{2}_{\bar{z}}}
\newcommand{\deby}{\partial_{\bar{y}}}
\newcommand{\debyq}{\partial^{2}_{\bar{y}}}

\begin{document}

\date{\mbox{}}

\title{
\vspace{-2.0cm}
\vspace{2.0cm}
{\bf \huge The critical tension in the Cascading DGP model}
 \\[8mm]
}

\author{
Fulvio Sbis\`a$^{1,2}$\thanks{fulvio.sbisa@port.ac.uk} \phantom{i}and Kazuya Koyama$^{1}$\thanks{kazuya.koyama@port.ac.uk}
\\[8mm]
\normalsize\it
$^1$ Institute of Cosmology \& Gravitation, University of Portsmouth,\\
\normalsize\it Dennis Sciama Building, Portsmouth, PO1 3FX, United Kingdom\vspace{.5cm} \\
\normalsize\it
$^2\,$ Dipartimento di Fisica dell'Universit\`a di Milano\\
       {\it Via Celoria 16, I-20133 Milano} \\
       and \\
\normalsize\it
       INFN, Sezione di Milano, \\
       {\it Via Celoria 16, I-20133 Milano}\vspace{.5cm} \\[.3em]
}

\maketitle

\setcounter{page}{1}
\thispagestyle{empty}

\begin{abstract}

\noindent We study the behaviour of weak gravitational fields in the 6D Cascading DGP model using a bulk-based approach. To deal with the ambiguity in the thin limit of branes of codimension higher than one, we consider a specific regularization of the internal structure of the branes where the 5D brane can be considered thin with respect to the 4D one. We consider the solutions corresponding to pure tension sources on the 4D brane, and study perturbations at first order around these background solutions. We adopt a 4D scalar-vector-tensor decomposition, and focus on the scalar sector of perturbations. We show that, in a suitable 4D limit, the trace part of the 4D metric perturbations obeys a decoupled equation which suggests that it is a ghost for background tensions smaller than a critical tension, while it is a healthy field otherwise. We give a geometrical interpretation of the existence of the critical tension and of the reason why the relevant field is a ghost or not depending on the background tension. We however find a value of the critical tension which is different from the one already found in the literature. Differently from the results in the literature, our analysis implies that, choosing the background tension suitably, we can construct ghost-free models for any value of the free parameters of the theory. We suggest that the difference lies in the procedure used to evaluate the pillbox integration across the codimension-2 brane. We confirm the validity of our analysis by performing numerically the integration in a particular case where the solution inside the thick cod-2 brane is known exactly. We stress that the singular structure of the perturbation fields in the nested branes set-ups is very subtle, and that great care has to be taken when deriving the codimension-2 junction conditions.

\end{abstract}

\smallskip

\section{Introduction}

Fifteen years after its discovery \cite{Riess98,Perlmutter98}, the problem of the cosmological late time acceleration remains puzzling. On one hand, recent cosmological observations are well fitted by $\La$CDM models. On the other hand, the best fit value for the cosmological constant $\La$ is dramatically different from the theoretical estimates for the value of vacuum energy \cite{WeinbergCC}, and yet different from zero. Along with explaining why $\La$ is small compared to the theoretical predictions (\emph{old} cosmological constant problem), it is now necessary to explain also why it is non-zero and extremely fine-tuned (\emph{new} cosmological constant problem). An intriguing interpretation of these observations is that they may signal a breakdown of General Relativity (GR) at ultra large scales, without the need of introducing an ad-hoc dark energy component.

From this point of view, a promising direction is to study theories which modify GR in the infrared, while reproducing its results at length scales where the latter is well tested (the \emph{modified gravity} approach). In the past years, several proposals have been developed following this idea, including $f(R)$ gravity, massive gravity and braneworld models (see \cite{ModifiedGravityAndCosmologyHugeReview} for a review). In particular, the  latter models (see \cite{Akama82,RubakovShaposhnikov83} for early proposals) are appealing from the point of view of high energy physics, since the existence of branes and of extra dimensions is an essential ingredient in string theory (see for example \cite{PolchinskiBook}). From the point of view of the cosmological constant problem, braneworld models with infinite volume extra dimensions can bypass Weinberg's no-go theorem \cite{ArkaniHamed:2002fu,Dvali:2002pe}, and more specifically (in the case of codimension higher than one) they may act as a high-pass filter on the wavelength of gravitational sources, effectively ``degravitating'' sources which are nearly constant with respect to a characteristic length of the model \cite{Dvali:2007kt}. From the point of view of the late time acceleration of the universe, the inclusion of an induced gravity term in the brane action (pioneered by the DGP model \cite{DGP00}) generically allows the existence of self-accelerating cosmological solutions, which may be used to explain the late time acceleration in a geometrical way.

The DGP model, however, does not provide a viable resolution of the acceleration problem, since its self-accelerating solutions are observationally ruled out \cite{MaartensMajerotto06,Rydbeck:2007gy,Fang:2008kc}, and are plagued by the presence of ghosts \cite{Koyama:2005tx, Gorbunov:2005zk, LutyPorratiRattazzi, NicolisRattazzi, DGPspectereoscopy, Koyama:2007za}. A natural idea is to consider higher codimension generalizations of the DGP model, since increasing the codimension should soften the tension between data and cosmological solutions \cite{Agarwal:2009gy}, and these models may provide a realization of the degravitation mechanism. However, increasing the codimension seems not to help with the ghost problem \cite{Dubovsky:2002jm} (however see \cite{Kolanovic:2003am,Berkhahn:2012wg}). Moreover, branes of codimension higher than one suffer from the notorious shortcomings that the thin limit of a brane is not well-defined \cite{GerochTraschen}, and that the brane-to-brane propagator of the gravitational field diverges when we send to zero the thickness of the brane \cite{Cline:2003ak, Vinet:2004bk} (unless we allow for Gauss-Bonnet terms in the bulk action \cite{Bostock:2003cv}). An important direction to explore is to consider elaborate constructions with more than one brane (as for example intersecting brane scenarios \cite{Corradini:2007cz, Corradini:2008tu}), hoping that the interplay between the branes may provide a mechanism to get rid of the ghosts. In particular, the Cascading DGP model \cite{deRham:2007xp} seems particularly interesting since it has been claimed that the gravitational field remains finite everywhere in the thin limit of the codimension-2 brane \cite{deRham:2007xp,deRham:2007rw} (phenomenon of \emph{gravity regularization}). Moreover, it has been shown that in the minimal (6D) formulation of the model there exists a critical value $\blac$ for the tension of the codimension-2 brane such that first order perturbations around pure tension backgrounds contain a ghost mode or not depending on the background tension $\bla$ \cite{deRham:2007xp,deRham:2010rw}. 

The purpose of this paper is to understand geometrically the mechanism which is responsible for the existence of the critical tension in the 6D Cascading DGP model. To deal with the ambiguity associated to the thin limit, we consider a specific realization of the system where the codimension-1 brane can be considered thin with respect to the codimension-2 brane. To study the behaviour of gravity we use a bulk-based point of view, which is geometrically more suited to the characteristics of the model. More precisely, we consider (background) configurations where the source on the codimension-2 brane has the form of pure tension, and we study the behaviour of the gravitational field at first order in perturbations around these solutions. We perform a 4D scalar-vector-tensor decomposition and focus on the scalar sector, since the analysis of \cite{deRham:2010rw} suggests that this is the only sector which is relevant for the existence of the critical tension.

The paper is structured as follows: in section \ref{Nested branes realization of the Cascading DGP model} we introduce our choice regarding the regularization of the internal structures of the branes, and we describe in detail the set-up. In section \ref{Scalar perturbations around pure tension solutions} we study scalar perturbations at first order around the pure tension solutions. In section \ref{The critical tension and ghost-free regions in parameters space} we derive the critical tension and we give a geometric interpretation of its existence. We furthermore discuss the difference between our result and the one in the literature, and we support our analysis with a numerical check. We finally present some conclusions in section \ref{Conclusions}.

\textbf{Conventions}: For metric signature, connection, covariant derivative, curvature tensors and Lie derivative we follow the conventions of Misner, Thorne and Wheeler \cite{MisnerThorneWheeler}. The metric signature is the ``mostly plus'' one, and we def\mbox{}ine symmetrization and antisymmetrization \emph{without} normalization. 6D indices are denoted by capital letters, so run from 0 to 5; 5D indices are denoted by latin letters, and run from 0 to 4, while 4D indices are denoted by greek letters and run from 0 to 3. The only exception is that the letters $i$, $j$ and $k$ indicate 2D indices which run on the extra dimensions $z$ and $y$. In general, quantities pertaining to the cod-1 brane are denoted by a tilde $\tilde{\phantom{a}}$, while quantities pertaining to the cod-2 brane are denoted by a superscript $\phantom{a}^{_{(4)}}$. Abstract tensors are indicated with bold-face letters, while quantities which have more than one component but are not tensors (such as coordinates $n$-tuples for example) are expressed in an abstract way replacing every index with a dot. When studying perturbations, the symbol $\simeq$ indicates usually that an equality holds at linear order. We use throughout the text the (Einstein) convention of implicit summation on repeated indices, and we will use unit of measure where the speed of light has unitary value $c=1$.

\section{Nested branes realization of the Cascading DGP model}
\label{Nested branes realization of the Cascading DGP model}

The Cascading DGP model \cite{deRham:2007xp} is a braneworld model where a $N$-dimensional bulk contains a recursive embedding of branes, each one equipped with an induced gravity term. In its minimal formulation, a 4D brane (which ought to describe our universe) is embedded in a 5D brane which is in turn embedded in a 6D bulk, and is qualitatively described by the action
\beq
\label{CascadingDGP6D}
S = \Msf \int_{\mcal{B}} \!\! d^6 X \, \sqrt{-g} \, R + \Mft \int_{\mcal{C}_1} \!\! d^5 \xi \, \sqrt{-\ti{g}} \, \ti{R} + \int_{\mcal{C}_2} \!\! d^4 \ch \, \sqrt{-g^{(4)}} \, \Big( \Mfs R^{(4)} + \mscr{L}_{M} \Big)
\eeq
where $\mcal{B}$ indicates the bulk, $\mcal{C}_1$ indicates the cod-1 brane and $\mcal{C}_2$ indicates the cod-2 brane. Here $\mathbf{g}$ indicates the bulk metric, $\ti{\mathbf{g}}$ indicates the metric induced on the cod-1 brane, while $\mathbf{g}^{(4)}$ indicates the metric induced on the cod-2 brane and the Lagrangian $\mscr{L}_{M}$ describes the matter localized on the cod-2 brane. Concerning the coordinate systems, the bulk is parametrized by the coordinates $\Xd = (z,y,x^{\m})$, the cod-1 brane is parametrized by the coordinates $\xid = (\xi, \xi^{\m})$ and the cod-2 brane is parametrized by the coordinates $\chd$. Similarly to the DGP model, a $Z_2$ reflection symmetry is enforced across the cod-1 brane; in addition to that, in the original formulation \cite{deRham:2007xp}
another $Z_2$ reflection symmetry is imposed in the bulk in the ``parallel'' direction to the cod-1 brane, so that in total the bulk enjoys a $Z_2 \times Z_2$ (double) reflection symmetry (by continuity, a $Z_2$ symmetry is imposed on the cod-1 brane). The theory has two free parameters, and it is convenient to use the mass scales
\begin{align}
\label{6DCascadingMassScales}
m_5 &\equiv \frac{\Mft}{\Mfs} & m_6 &\equiv \frac{\Msf}{\Mft}
\end{align}
and the associated length scales $l_5 \equiv 1/m_5$ and $l_6 \equiv 1/m_6$. The analysis of \cite{deRham:2007xp,deRham:2007rw} shows that gravity behaves in a qualitatively different way depending on the relation between $m_5$ and $m_6$: if $m_5 \gg m_6$, weak gravity ``cascades'' from a 6D behaviour at very large scales to a 5D behaviour at intermediate scales to a 4D behaviour at small scales, while if $m_5 \ll m_6$ there is a direct transition from a 6D behaviour at large scales to a 4D behaviour at small scales. As in the DGP model, at small scales the tensor structure of the weak gravitational field is different form GR's, so at linear level the theory does not reproduce GR results. However, it is expected that the agreement with GR is recovered at non-linear level \cite{deRham:2007rw} via a multiple Vainshtein mechanism \cite{Vainshtein72,Deffayet:2001uk}. See \cite{deRham:2010rw, Minamitsuji:2008fz, Khoury:2009tk, deRham:2009wb, Agarwal:2009gy, Wyman:2010jp, Agarwal:2011mg, Moyassari:2011nb, Rug:2012tx, Hao:2014tsa} for other studies related to the Cascading DGP model.

\subsection{The nested branes realization }
\label{The nested branes realization}

It is crucial to notice that the action (\ref{CascadingDGP6D}) a priori does not single out a unique model if the internal structures of the branes $\mcal{C}_1$ and $\mcal{C}_2$ are not specified. In fact, it is well-known that the thin limit of branes of codimension higher than one is not well defined \cite{GerochTraschen}: it is expected that this property does not change if we embed a cod-2 brane inside a cod-1 brane, since, beside the freedom to choose the cod-2 internal structure, we now have the additional freedom to choose how the internal structures of the two (cod-1 and cod-2) branes are related one to the other (see \cite{Sbisa':2014uza} for a discussion on this point). In absence of a rigorous proof (on the lines of \cite{GerochTraschen}) of the well-definiteness of the thin limit of the Cascading DGP model, to perform a transparent analysis we should explicitly take into account the internal structures of the branes and their mutual relationship.

An interesting choice in this sense is to consider configurations where the thickness of the cod-1 brane is much smaller than the thickness of the cod-2 brane, so that the former can be considered thin with respect to the latter. In fact, since the thin limit of a cod-1 brane is well-defined, we can describe these configurations as if the cod-1 brane were (infinitely) thin, and the cod-2 brane were ``ribbon'' inside the cod-1 brane. Moreover, the results of \cite{Sbisa':2014uza} imply that the thin limit of the ribbon brane is well-defined (at least at first order in perturbations), so fixing this hierarchy permits to work with thin branes. Therefore, in the following we consider only this class of configurations, to which we refer as the \emph{nested branes realization of the Cascading DGP model} (due to the close connection with the nested branes with induced gravity set-ups introduced in \cite{Sbisa':2014uza}).

\subsubsection{The set-up}

It is worthwhile to be more specific about what we mean when we say that a ``ribbon'' brane lies inside a cod-1 brane. First of all, we assume that a 5D submanifold $\mcal{C}_1$ (the cod-1 brane) is embedded in the 6D bulk, and we impose a $\mathbb{Z}_{2}$ symmetry across the cod-1 brane. Secondly, we assume that a 4D submanifold $\mcal{C}_2$ is embedded inside the cod-1 brane, and that matter, tension and a 4D induced gravity term are confined inside the brane $\mcal{C}_1$ and are localized around the brane $\mcal{C}_2$. We impose that a $\mathbb{Z}_{2}$ symmetry across $\mcal{C}_2$ holds inside the cod-1 brane. More specifically, we distinguish between a physical (thick) cod-2 brane, inside which energy and momentum are confined (the ``ribbon'' cod-2 brane), and a mathematical (thin) cod-2 brane ($\mcal{C}_2$), with respect to which the $\mathbb{Z}_{2}$ symmetry is imposed. When the thin limit of the cod-2 brane is performed, the physical brane coincides with the mathematical one. Note that, differently from the original formulation of the Cascading DGP model \cite{deRham:2007xp}, we do \emph{not} impose a $\mathbb{Z}_{2} \times \mathbb{Z}_{2}$ symmetry to hold in each of the two mirror copies which constitute the bulk. In fact, the presence of a $\mathbb{Z}_{2}$ symmetry inside the cod-1 brane does not imply that a double $\mathbb{Z}_{2}$ symmetry holds outside of it.

Following the conventions of \cite{Sbisa':2014uza}, we describe the position of the cod-1 brane in the bulk by the embedding function $\vfd$ whose component expression is $\varphi^{A}(\xi^{a})$, while we describe the position of the (mathematical) cod-2 brane inside the cod-1 brane by the embedding function $\ti{\a}^{\cdot}$ whose expression in coordinates is $\ti{\a}^{a}(\chi^{\m})\,$. Composing these two embedding functions, we obtain the embedding function $\b^{\cdot}$ of the mathematical cod-2 brane in the bulk which in coordinated reads $\b^{A}(\ch^{\m}) = \varphi^{A} \big( \ti{\a}^{a}(\ch^{\m}) \big)$. The bulk metric $\mathbf{g}$ induces on the codimension-1 brane the metric $\ti{\mathbf{g}} \equiv \varphi_{\star} \big( \mathbf{g} \big)$, where $\varphi_{\star}$ indicates the pullback with respect to $\vf^{\cdot}$, and in turn the metric $\ti{\mathbf{g}}$ induces on the codimension-2 brane a metric $\mathbf{g}^{_{(4)}} \equiv \ti{\a}_{\star} \big( \mathbf{\ti{g}} \big)$, where $\ti{\a}_{\star}$ indicates the pullback with respect to $\ti{\a}^{\cdot}$. The solution of equations of motion for this set-up are found by solving the Einstein equations in the bulk
\beq
\label{Bulkeq}
\mbfG = \,\, 0
\eeq
and by imposing the Israel junction conditions \cite{Israeljc} at the cod-1 brane
\beq
\Msf \Big[ \tmbfK - \ti{K} \, \tmbfg \Big]_{\pm} + \Mft \, \tmbfG = \,\, \tmbfT
\eeq
where $\tmbfK$ is the extrinsic curvature\footnote{Following \cite{Sbisa':2014uza}, we indicate with $\tmbfK$ the pullback on the cod-1 brane of the second fundamental form.} of the cod-1 brane ($\ti{K}$ is its trace), $\tmbfG$ is the Einstein tensor built from the cod-1 induced metric and $\tmbfT$ is the generalized energy-momentum tensor on the ribbon cod-2 brane. Taking advantage of the $\mathbb{Z}_2$ symmetry which holds across the cod-1 brane, it is enough to solve the Einstein equations only in one of the two mirror copies which constitute the bulk (henceforth, with a slight abuse of language we refer to the chosen copy as the ``bulk'' itself), and to impose the junction conditions at the boundary
\beq
\label{junctionconditionseq}
2 \Msf \Big( \tmbfK - \ti{K} \, \tmbfg \Big) + \Mft \, \tmbfG = \,\, \tmbfT
\eeq

The mirror symmetry present inside the cod-1 brane is explicitly realized when we use coordinate systems on the cod-1 brane which are Gaussian Normal with respect to the brane $\mcal{C}_2$. Henceforth, we refer to this class of reference systems as codimension-1 Gaussian Normal reference systems (or briefly cod-1 GNC), and we indicate quantities evaluated in this coordinate systems with an overhat $\h{\phantom{a}}$. These reference systems are always well-defined at least in a neighbourhood of $\mcal{C}_2$, and are constructed starting from a reference system $\chd$ on the cod-2 brane and following the geodesic of $\mcal{C}_1$ which are normal to $\mcal{C}_2$. We synthetically indicate the cod-1 GN coordinates as $\hxid \equiv ( \hxi, \chd )$, and by construction we have that \cite{CarrollBook}
\begin{align}
\label{c1GNCgeneral}
\h{g}_{\xi\xi}(\hxi,\chd) &= 1 & \h{g}_{\xi\m}(\hxi,\chd) &= 0
\end{align}
The requirement that a $\mathbb{Z}_{2}$ symmetry across the cod-2 brane holds inside the cod-1 brane implies that, when expressed in cod-1 GNC, the $\m\n$ and $\xi\xi$ components of the induced metric $\h{\mathbf{g}}$ (as well as of the tensors $\h{\mathbf{G}}$, $\h{\mathbf{K}}$ and $\h{\mathbf{T}}$) are symmetric with respect to the reflection $\hxi \rightarrow - \hxi$, while the $\xi\m$ components are antisymmetric.

\subsubsection{The structure of the source term}

Concerning the source term on the cod-1 brane, we assume that the generalized energy-momentum tensor $\ti{\mathbf{T}}$ present on the cod-1 brane is localized around the cod-2 brane $\mcalC_2$. By this we mean that there exists a (finite) localization length $l_2$ such that in cod-1 GNC the tensor $\h{T}_{ab}(\hxi, \chd)$ vanishes when it is evaluated at a distance $\hxi$ from the cod-2 brane which is bigger than $l_2$ (so the thickness of the ribbon brane is $2 \, l_2$). More specifically, we assume the following structure for the generalized energy-momentum tensor
\beq
\label{marina}
\h{T}_{ab}(\hxi,\chd) = - f_1(\hxi) \, \la \,\, \d_{a}^{\, \, \m} \,\, \d_{b}^{\, \, \n} \,\, \g_{\m\n}(\hxi;\chd) + \h{\mcalT}_{ab}(\hxi,\chd) - f_2(\hxi) \,\, \d_{a}^{\, \, \m} \,\, \d_{b}^{\, \, \n} \,\, \Mfs \, \mcal{G}_{\m\n}(\hxi;\chd)
\eeq
where $\h{\mcalT}_{ab}$, $f_1$ and $f_2$ vanish for $\abs{\hxi} > l_2$. In the last expression, $\g_{\m\n}(\hxi;\chd)$ is a one-parameter family of (tensor) functions of the 4D coordinates $\chd$ which coincide with the 4D components of the induced metric $\h{g}_{ab}$ when the latter is expressed in cod-1 GNC
\beq
\g_{\m\n}(\hxi;\chd) \equiv \Big[ \h{\mathbf{g}}(\hxi,\chd) \Big]_{\m\n}
\eeq
The one-parameter family of tensors $\mcal{G}_{\m\n}$ is defined such that, for every value of $\hxi$, $\mcal{G}_{\m\n}(\hxi;\chd)$ is the 4D Einstein tensor built from the metric $\g_{\m\n}(\hxi;\chd)$. The \emph{localizing functions} $f_1$ and $f_2$ are even functions (to comply with the $\mathbb{Z}_2$ symmetry) which are regularized versions of the Dirac delta function, i.e.~they obey
\beq
\label{Samatorza}
\int_{-l_2}^{+l_2} \! d\hxi \,\, f_1 (\hxi) \, = \int_{-l_2}^{+l_2} \! d\hxi \,\, f_2 (\hxi) \, = 1
\eeq

Let us comment on each of the three contributions which constitute the generalized energy-momentum tensor. Note that $\g_{\m\n}(\z;\chd)$ is the 4D metric induced on the $\hxi$--constant (4D) hypersurface defined by $\hxi = \z$. This implies that, on every $\hxi$--constant hypersurface, the term $- f_1 \, \la \, \d_{a}^{\, \, \m} \, \d_{b}^{\, \, \n} \, \g_{\m\n}$ has the form of pure tension, where the total tension $\la$ is distributed in the $\hxi$ direction according to the function $f_1$. In the thin limit (where $f_1$ and $f_2$ tend to the Dirac delta), this terms tends to
\beq
- f_1(\hxi) \, \la \,\, \d_{a}^{\, \, \m} \,\, \d_{b}^{\, \, \n} \,\, \g_{\m\n}(\hxi;\chd) \quad \to \quad - \d(\hxi) \, \la \,\, \d_{a}^{\, \, \m} \,\, \d_{b}^{\, \, \n} \,\, g^{(4)}_{\m\n}(\chd)
\eeq
which is the energy-momentum tensor correspondent to having pure tension $\la$ on the thin cod-2 brane. Therefore, the first term in the right hand side of (\ref{marina}) describes a thick pure tension source on the ribbon brane. The term $\h{\mcalT}_{ab}(\hxi,\chd)$ instead is the energy-momentum tensor of matter present inside the ribbon brane. To formalize the idea that momentum does not flow out of the brane, we ask that the pillbox integration of the normal and mixed components of $\h{\mcalT}_{ab}$ vanishes
\beq
\label{angelica}
\int_{-l_2}^{+l_2} d \hxi \,\, \h{\mcalT}_{\xi\xi}(\hxi, \chd) = \int_{-l_2}^{+l_2} d \hxi \,\, \h{\mcalT}_{\xi\m}(\hxi, \chd) = 0
\eeq
and we define the cod-2 matter energy-momentum tensor as the 4D tensor $\mcalT^{_{(4)}}_{\m\n}(\chd)$ obtained by the pillbox integration of the 4D components of $\h{T}_{ab}$
\beq
\label{cristina}
\int_{-l_2}^{+l_2} d \hxi \,\, \hat{\mcalT}_{ab}(\hxi, \chd) = \d_{a}^{\, \, \m} \,\, \d_{b}^{\, \, \n} \,\, \mcalT^{(4)}_{\m\n}(\chd)
\eeq
The term $- f_2(\hxi) \, \d_{a}^{\, \, \m} \, \d_{b}^{\, \, \n} \, \Mfs \, \mcal{G}_{\m\n}(\hxi;\chd)$ is instead a 4D induced gravity term which, instead of being localized at $\hxi = 0$, is distributed in the $\hxi$ direction according to the function $f_2$. In the thin limit this term tends to
\beq
- f_2(\hxi) \,\, \d_{a}^{\, \, \m} \,\, \d_{b}^{\, \, \n} \,\, \Mfs \, \mcal{G}_{\m\n}(\hxi;\chd) \quad \to \quad - \d(\hxi) \,\, \d_{a}^{\, \, \m} \,\, \d_{b}^{\, \, \n} \,\, \Mfs \, G^{(4)}_{\m\n}(\chd)
\eeq
which is the induced gravity term for the thin cod-2 brane. To conclude, the generalized energy-momentum tensor (\ref{marina}) corresponds to a configuration where matter, tension and a 4D induced gravity term are distributed on a thick ribbon brane (see \cite{deRham:2007rw} for a discussion of the gauge invariance properties of the procedure of smoothing a localized action).

\section{Scalar perturbations around pure tension solutions}
\label{Scalar perturbations around pure tension solutions}

We now study the behaviour of weak gravitational fields in the nested branes realization of the 6D Cascading DGP model. More precisely, we consider pure tension (background) configurations, and study perturbations at first order around these configurations. We consider a 4D scalar-vector-tensor decomposition of the perturbation modes, and focus on the scalar sector, which is relevant for the critical tension. We indicate the quantities correspondent to the background configurations with an overbar $\bar{\phantom{i}}$.

\subsection{Pure tension solutions}
\label{Pure tension solutions}

Let's consider localized source configurations where matter is absent $\ti{\mcalT}_{ab} = 0$, so in cod-1 Gaussian Normal Coordinates the generalized energy-momentum tensor is of the form
\beq
\label{marinaB}
\bar{T}_{ab}(\hxi, \chd) = - f_1(\hxi) \, \bla \,\, \d_{a}^{\, \, \m} \,\, \d_{b}^{\, \, \n} \,\, \g_{\m\n}(\hxi;\chd) - f_2(\hxi) \,\, \d_{a}^{\, \, \m} \,\, \d_{b}^{\, \, \n} \,\, \Mfs \, \bar{\mcal{G}}_{\m\n}(\hxi;\chd)
\eeq
where $\bla > 0$. It has been shown in \cite{Sbisa':2014uza} (building on the previous works \cite{Dvali:2006if,Gregory:2001xu,Gregory:2001dn}) that when $\Mfs = 0$ there exist solutions where the (mathematical) cod-2 brane $\mathcal{C}_2$ is placed at $\xi = 0$
\beq
\label{Robert}
\bar{\a}^{a}(\chi^{\cdot}) = \big( 0, \chi^{\m} \big)
\eeq
and the bulk is flat
\beq
\label{George}
\bar{g}_{AB}(X^{\cdot}) = \eta_{AB}
\eeq
while the cod-1 brane has the following embedding
\beq
\label{Patty}
\bar{\varphi}^{A}(\xi^{\cdot}) = \big( Z(\xi), Y(\xi), \xi^{\m} \big)
\eeq
where
\begin{align}
\label{Jerry}
\Zp\big( \hxi \big) &= \sin S \big( \hxi \big)  & \Yp\big( \hxi \big) &= \cos S \big( \hxi \big) 
\end{align}
and the ``slope function'' $S$ reads
\beq
\label{satisfaction}
S(\hxi) = \frac{\bla}{2 \Msf} \, \int_{0}^{\hxi} f_1 \big( \z \big) d \z
\eeq
Since in this case the metric $\g_{\m\n}(\hxi;\chd)$ is the 4D Minkowski metric, these configurations are solutions of the equations of motion also when $\Mfs \neq 0$. We can freely impose the conditions $Z(0) = Y(0) = 0$ so that $Z$ is even with respect to the parity transformation $\hxi \to - \hxi$ while $Y$ is odd.

It is easy to see \cite{Sbisa':2014uza} that, outside the thick cod-2 brane, the slope function is constant
\beq
S(\hxi) = \pm S_+ \equiv \frac{\bla}{4 \Msf} \quad \qquad \textrm{for} \quad \qquad \hxi \gtrless \pm l_2
\eeq
and is determined only by the total amount of tension $\bla$ (it is independent of how the tension is distributed inside the brane). Therefore, the thin limit of these solutions exists and is given by the configurations where the tension $\bla$ is perfectly localized on $\mcal{C}_2$ and the components of the embedding function read
\begin{align}
\label{Iris}
Z(\hxi) &= \sin \Big( \frac{\bla}{4 M_6^4} \Big) \,\, \abs{\hxi} & Y(\hxi) &= \cos \Big( \frac{\bla}{4 M_6^4} \Big) \,\, \hxi
\end{align}
Note that the normal 1-form reads\footnote{As we discuss in \cite{Sbisa':2014uza}, this is the choice of the normal form with the correct orientation.}
\beq
\label{Irioth}
\bn_A(\hxi) = \big( \Yp(\hxi) , - \Zp(\hxi) , 0,0,0,0 \big)
\eeq
and becomes discontinuous in the thin limit. The complete 6D spacetime which corresponds to these thin limit configurations is the product of the 4D Minkowski space and a two dimensional cone of deficit angle $\a = \bla/M_6^4$. When $\bla \to 2 \pi M_{6}^{4 \, -}$ the deficit angle tends to $2 \pi$, and the 2D cone tends to a degenerate cone (a half-line). Therefore there is an upper bound $\bla < \bla_M \equiv 2 \pi M_{6}^{4}$ on the tension which we can put on the thin cod-2 brane.

\subsection{Small perturbations in the bulk-based approach}
\label{Small perturbations in the bulk-based approach}

As we discuss in \cite{Sbisa':2014uza}, to find consistent solutions in the thin limit of the ribbon brane we need either embedding functions which are cuspy, or a bulk metric which is discontinuous, or both. This is necessary to produce a delta function divergence in the left hand side of the junction conditions (which balances the delta function divergence on the right hand side), while at the same time keeping the gravitational field on the thin cod-2 brane finite (gravity regularization).

The choice to privilege a smooth bulk metric, without constraining the form of the embedding, or to privilege a smooth embedding, leaving the bulk metric free to have discontinuities, is related to adopting a bulk-based or a brane-based point of view. We suggest in \cite{Sbisa':2014uza} that the bulk-based approach have several advantages. In fact, it permits to identify clearly the degrees of freedom which are responsible for the singularity (the embedding functions), separating them from the degrees of freedom which are not (the bulk metric). Moreover, the property of gravity being finite is mirrored by the fact that all the degrees of freedom remain continuous in the thin limit, so the regularity properties of the solutions are tightly linked to the regularity properties of the gravitational field. In addition, the global geometry of the thin limit configurations is more transparent in the bulk-based approach, for example in the pure tension case the deficit angle is directly connected to the slope of the embedding. As we shall see, there is also a more technical reason in favour of this choice: the bulk-based approach permits us to identify clearly the convergence properties of the perturbative degrees of freedom in the thin limit.

For these reasons, we adopt the bulk-based approach to study small perturbations around the pure tension solutions, following closely the analysis of \cite{Sbisa':2014uza}.

\subsubsection{Perturbations around pure tension solutions}

We consider the following perturbative decomposition for the bulk and bending degrees of freedom
\begin{align}
g_{AB}(X^{\cdot}) &= \bar{g}_{AB}(X^{\cdot}) + h_{AB}(X^{\cdot}) \\[2mm]
\varphi^{A}(\xi^{\cdot}) &= \bar{\varphi}^{A}(\xi^{\cdot}) + \d\!\varphi^{A}(\xi^{\cdot})
\end{align}
and do not constrain the form of $h_{AB}$ and $\d\!\varphi^{A}$, so we leave both the bulk metric and the cod-1 embedding free to fluctuate. We instead decide to keep fixed the position of the cod-2 brane inside the cod-1 brane, and we still use the 4D coordinates of the cod-1 brane to parametrize the cod-2 brane, so the embedding of the cod-2 brane in the cod-1 brane reads
\begin{align}
\label{Joan}
\tilde{\a}^{a}(\chi^{\cdot}) &= \bar{\a}^{a}(\chi^{\cdot}) = \big( 0, \chi^{\m} \big) & \d \tilde{\a}^{a}(\chi^{\cdot}) &= 0
\end{align}
also at perturbative level. However, the freedom of the cod-1 brane to fluctuate in the bulk implies that also the cod-2 brane is free to fluctuate in the bulk
\beq
\b^{A}(\chd) = \bar{\b}^{A}(\chd) + \d\!\b^{A}(\chd)
\eeq
where
\begin{align}
\bar{\b}^{A}(\chd) &= \bvf^{A} \big( 0, \chd \big) & \d\!\b^{A}(\chd) &= \d\!\varphi^{A} \big( 0, \chd \big)
\end{align}
We def\mbox{}ine the perturbations of the metric induced on the cod-1 brane as follows
\beq
\label{Caribbean}
\tilde{h}_{ab}(\xi^{\cdot}) \equiv  \tilde{g}_{ab}(\xi^{\cdot}) - \bar{g}_{ab}(\xi^{\cdot})
\eeq
and analogously we define the perturbation of the metric induced on the cod-2 brane as
\beq
h^{(4)}_{\m \n}(\chi^{\cdot}) \equiv g^{(4)}_{\m \n}(\chi^{\cdot}) - \bar{g}^{(4)}_{\m \n}(\chi^{\cdot})
\eeq

It is useful to introduce the vectors tangent to the cod-1 brane in the $\xi$ direction
\beq
v^{A} \equiv \frac{\de \vf^A}{\de \xi}
\eeq
which can be perturbatively decomposed as follows
\beq
v^{A} = \bv^{A} + \dv^{A}
\eeq
where
\begin{align}
\bv^{A} &= \big( \Zp, \Yp, 0, 0, 0, 0 \big) & \dv^{A} &= \dvf^{A \, \p}
\end{align}
and we indicated a derivative with respect to $\xi$ with a prime $\phantom{i}^{\p}$. We adopt the convention that indices on background quantities and on perturbations are lowered/raised with the background metric and its inverse, so for example $\bv_A = \e_{AL} \bv^L$. It follows that the 2D indices $i$, $j$ and $k$, which run on the extra dimensions $z$ and $y$, are raised/lowered with the identity matrix, so we have for example
\begin{align}
\bv_i = \bvf_i^{\p} &\equiv \d_{ij} \, \bvf^{j \, \p} & \bn^{i} &\equiv \d^{ij} \, \bn_{j}
\end{align}

Concerning the source term, we perturb both the matter content and the tension of the cod-2 brane, so in cod-1 GNC we have
\beq
\hat{T}_{ab} = \bar{T}_{ab} - \d_{a}^{\, \, \m} \,\, \d_{b}^{\, \, \n} \, f_1(\hxi) \, \bla \, \h{h}_{\m\n} + \d \hat{T}_{ab}
\eeq
where $\bar{T}_{ab}$ is the background source term
\beq
\bar{T}_{ab} = - \d_{a}^{\, \, \m} \,\, \d_{b}^{\, \, \n} \, f_1(\hxi) \, \bla \,\, \e_{\m\n}
\eeq
and 
\beq
\label{deltaTmunu}
\d \hat{T}_{ab}(\hxi, \chd) = - \d_{a}^{\, \, \m} \,\, \d_{b}^{\, \, \n} \, f_1(\hxi) \,\, \dla \,\, \e_{\m\n} + \h{\mcalT}_{ab}(\hxi,\chd) - f_2(\hxi) \,\, \d_{a}^{\, \, \m} \,\, \d_{b}^{\, \, \n} \,\, \Mfs \, \mcal{G}_{\m\n}(\hxi;\chd)
\eeq
where $\dla$ is the perturbation of the tension.

\subsubsection{The 4D scalar-vector-tensor decomposition}

To deal with the issue of gauge invariance, we perform a 4D scalar-vector-tensor decomposition of the bulk and brane degrees of freedom and work with gauge invariant variables. We consider in fact the following decomposition of the bulk metric perturbations
\begin{align}
h_{\mu \nu} &= \mscr{H}_{\mu \nu} + \partial_{(\mu} V_{\nu)} + \e_{\mu \nu} \,\pi + \partial_{\mu}\partial_{\nu} \vp \\[2mm]
h_{z \mu} &= A_{z \mu} + \partial_{\mu} \s_{z} \\[2mm]
h_{y \mu} &= A_{y \mu} + \partial_{\mu} \s_{y} \\[2mm]
h_{yy} &= \psi \\[2mm]
h_{zy} &= \r \\[2mm]
h_{zz} &= \o
\end{align}
where all these quantities are functions of the bulk coordinates $X^{\cdot}$, and we used the notation $\dem \equiv \de/\de x^\m$. In particular, $\mscr{H}_{\mu \nu}$ is a transverse-traceless symmetric tensor while $V_\mu$ , $A_{z \mu}$ and $A_{y \mu}$ are transverse 1-forms, and $\o$, $\r$, $\psi$, $\s_{z}$, $\s_{y}$, $\pi$ and $\vp$ are scalars.

Concerning the codimension-1 brane, we consider the scalar-vector-tensor decomposition with respect to the 4D coordinates $\xim$. Regarding the embedding, the bending modes $\dvf^z$ and $\dvf^y$ are scalars, while the 4D components can be decomposed as
\beq
\dvf_{\m} = \dvf^{_{T}}_{\m} + \dexim \dvf_4
\eeq
where $\dvf_4$ is a scalar and $\dvf^{_{T}}_{\m}$ is a transverse vector, and $\dexim \equiv \de/\de \xi^\m$. Regarding the cod-1 induced metric, its decomposition is naturally linked to the decomposition of the bulk metric since each sector (scalar/vector/tensor) of the induced metric contains only the bulk perturbations of the corresponding sector \cite{Sbisa':2014uza}. This is true in turn for the decomposition with respect of the coordinates $\chd$ of the (double) induced metric on the mathematical cod-2 brane $\mcalC_2$. To avoid a cumbersome notation, we use henceforth the convention that the evaluation on the cod-1 brane of a bulk quantity is indicated with a tilde (or with an overhat if we are using cod-1 GNC), so for example
\begin{align}
\ti{\pi}(\xid) &\equiv \pi(\Xd) \evb & \hat{\pi}(\hxi, \chd) &\equiv \pi(\Xd) \evbh
\end{align}
and we use the convention that the evaluation on the (mathematical) cod-2 brane of a bulk quantity is indicated with a superscript $\!\phantom{i}^{_{(4)}}$, so for example
\beq
\pi^{(4)}(\chd) \equiv \pi(\Xd) \Big\rvert_{X^{\cdot} = \bar{\varphi}^{\cdot}(0,\chd)}
\eeq

Regarding the matter cod-1 energy-momentum tensor, we consider the following decomposition 
\begin{align}
\ti{\mcalT}_{\m\n} &= \ti{\mscrT}_{\m\n} + \de_{\xi^{( \m}} \, \ti{B}_{\n)} + \deximxin \, \ti{\mcalT}_{de} + \eta_{\m\n} \, \ti{\mcalT}_{tr}  \\[2mm]
\ti{\mcalT}_{\xi \mu} &= \ti{D}_{\mu} + \partial_{\xim} \tt
\end{align}
where the symmetric tensor $\ti{\mscrT}_{\m\n}$ is transverse and traceless, while $\ti{B}_{\m}$ and $\ti{D}_{\m}$ are transverse 1-forms and $\ti{\mcalT}_{\xi \xi}$, $\ti{\mcalT}_{tr}$, $\ti{\mcalT}_{de}$ are scalars. We consider also the scalar-vector-tensor decomposition of the cod-2 energy-momentum tensor with respect to the coordinates $\chm$
\beq
\mcalT^{(4)}_{\m\n} = \mscrT_{\m\n}^{(4)} + \de_{\ch^{( \m}} \, B^{(4)}_{\n)} + \dechmchn \, \mcalT^{(4)}_{de} + \eta_{\m\n} \, \mcalT^{(4)}_{tr}
\eeq
where $\mscrT_{\m\n}^{_{(4)}}$, $B^{_{(4)}}_{\m}$, $\mcalT^{_{(4)}}_{de}$ and $\mcalT^{_{(4)}}_{tr}$ are respectively the pillbox integration of $\h{\mscrT}_{\m\n}$, $\h{B}_{\m}$, $\h{\mcalT}_{de}$ and $\h{\mcalT}_{tr}$, while the pillbox integration of $\h{D}_{\mu}$, $\h{\t}$ and $\h{\mcalT}_{\xi\xi}$ vanish as a consequence of (\ref{angelica}). Note that the covariant conservation of the cod-2 energy momentum tensor implies
\beq
\boxf \mcalT^{(4)}_{de} + \mcalT^{(4)}_{tr} = 0 \label{nottoobad1}
\eeq
which in particular permits to express $\mcalT^{(4)}_{tr}$ and $\boxf \mcalT^{(4)}_{de}$ in terms of the trace $\mcal{T}^{(4)} \equiv \e^{\m\n} \, \mcal{T}^{(4)}_{\m\n}$ of the matter cod-2 energy-momentum tensor, namely
\begin{align}
\label{cod2enmomtrace}
\mcalT^{(4)}_{tr} &= \third \, \mcalT^{(4)} & \boxf \mcalT^{(4)}_{de} &= - \third \, \mcalT^{(4)}
\end{align}

\subsubsection{Gauge-invariant and master variables}

From now on we focus only on the scalar sector. As we showed in \cite{Sbisa':2014uza}, it is possible to describe in a gauge invariant way both the fluctuation of the bulk metric and the fluctuation of the cod-1 brane position. In particular, in the scalar sector there are four ``metric'' gauge invariant variables
\begin{align}
\pi^{gi} &\equiv \pi \\[2mm]
h_{ij}^{gi} &\equiv h_{ij} - \ded{(i} \, \s_{j)} + \de_{i} \de_{j} \, \varpi
\end{align}
where $h_{ij}^{gi}$ synthetically indicates $h_{zz}^{gi}$, $h_{zy}^{gi}$ and $h_{yy}^{gi}$, and three ``brane'' gauge invariant variables
\begin{align}
\dvf^{i}_{gi} &\equiv \dvf^{i} + \Big[ \s_i - \half \, \de_i \vp \Big]\evb \\[1mm]
\dvf^{gi}_{4} &\equiv \dvf_{4} + \half \, \tvp
\end{align}
where $\dvf^{i}_{^{gi}}$ synthetically indicates $\dvf^{z}_{^{gi}}$, $\dvf^{y}_{^{gi}}$. In addition, we can describe in a gauge-invariant way also the fluctuation of the position of the (mathematical) cod-2 brane in the extra dimensions, introducing the cod-2 gauge invariant bending modes
\beq
\d\!\b^{i}_{gi} \equiv \d\!\b^{i} + \Big[ \s_i - \half \, \de_i \vp \Big]\Big\rvert_{X^{\cdot} = \bar{\b}^{\cdot}(\chd)}
\eeq
The $\mathbb{Z}_2$ symmetry present inside the cod-1 brane however implies that $\d \b^{y}_{gi}$ vanish identically, so the relevant gauge invariant mode which describes the movement of the cod-2 brane in the bulk is the field
\beq
\d\!\b^{(4)} (\chd) = \d\!\b^{z}_{gi} (\chd)
\eeq

In the $\Mfs = 0$ case it is moreover possible \cite{Sbisa':2014uza} to express the equations in terms of master variables \cite{Mukohyama:2000ui,Kodama:2000fa}. In fact, the bulk equations imply that the gauge invariant variables $h_{ij}^{gi}$ can be expressed in terms of $\pi$ as follows 
\beq
\label{desiree2}
h^{gi}_{ij} = - 3 \, \d_{ij} \, \pi - \frac{4}{\boxf} \, \de_{i} \de_{j} \, \pi
\eeq
and so the metric part of the scalar sector can be expressed in terms of the master variable $\pi$ whose bulk equation is
\beq
\Box_6 \, \pi = 0 \label{Marcus} 
\eeq
For the brane part, it is convenient to define the normal and parallel component of the bending $\dvfn$ and the parallel component $\dvfp$
\begin{align}
\label{whoknows}
\dvfn &\equiv \bn_i \, \dvf_{gi}^{i} & \dvfp &\equiv \bv_i \, \dvf_{gi}^i
\end{align}
Since $\dvf^{gi}_{4}$ does not appear in the equations of motion, and $\dvfp$ does not appear in the thin limit \cite{Sbisa':2014uza}, the normal component of the bending $\dvfn$ is the master variable for the brane perturbations in the scalar sector (in the thin limit). It is important to keep in mind that, despite the normal and parallel components of the bending have an intuitive geometrical meaning when the normal vector is smooth, they are not well defined when the normal vector is discontinuous, while the $z$ and $y$ components of the bending remain well-defined.

These results do not change if we add the 4D induced gravity term on the ribbon brane, since we have explicitly
\beq
\mcal{G}_{\m\n}(\hxi;\chd) = - \half \, \boxf \, \h{\mscr{H}}_{\m\n}(\hxi,\chd) + \e_{\m\n} \, \boxf \, \hp(\hxi,\chd) - \dechmchn \, \hp(\hxi,\chd)
\eeq
and so the scalar sector of $\mcal{G}_{\m\n}$ can be expressed purely in terms of $\hp$. Therefore, $\pi$ and $\dvfn$ are the (scalar) master variables of the system also in the $\Mfs \neq 0$ case.

\subsection{Thin limit equations of motion}
\label{Thin limit equations of motion}

We turn now to the equations of motion for the perturbations. As we explained in \cite{Sbisa':2014uza}, when we take the thin limit of the ribbon brane the cod-1 junction conditions (\ref{junctionconditionseq}) split into two sets of equations: the pure codimension-1 junction conditions, which are source-free and hold for $\hxi \neq 0$, and the codimension-2 junction conditions, which are sourced and link the value of the solution at $\hxi = 0^-$ and $\hxi = 0^+$. The addition of the 4D induced gravity term $- f_2(\hxi) \, \d_{a}^{\, \, \m} \, \d_{b}^{\, \, \n} \, \Mfs \, \mcal{G}_{\m\n}(\hxi;\chd)$ to the cod-1 energy-momentum tensor does not spoil the derivation of the codimension-2 junction conditions, which proceeds exactly in the same way as in \cite{Sbisa':2014uza}. Therefore, the thin limit equations of motion for the nested branes realization of the Cascading DGP model are obtained simply by performing the substitution
\beq
\label{magicsubstitution}
\mcalT^{(4)}_{\m\n}(\chd)  \to \mcalT^{(4)}_{\m\n}(\chd) - \Mfs \, G^{(4)}_{\m\n}(\chd) 
\eeq
in the equations derived in \cite{Sbisa':2014uza}. Furthermore, in the $\Mfs = 0$ case a pure tension perturbation produces a perturbation of the deficit angle $\d \a = \dla/M_6^4$ while leaves flat the bulk metric \cite{Sbisa':2014uza}. Since the induced metric on the cod-2 brane remains flat as well, these solutions are valid also in the $\Mfs \neq 0$ case. Note that at linear order in perturbations the effect of a pure tension perturbation and of a matter perturbation are additive, so for simplicity henceforth we consider only the pure matter perturbation case $\dla = 0$ (with the exception of section \ref{Numerical check} and of appendix \ref{Pure tension perturbations}).

\subsubsection{Thin limit of the scalar sector}

Performing the substitution (\ref{magicsubstitution}) in the equations derived in \cite{Sbisa':2014uza}, and considering only the scalar sector, the bulk equations of motion in term of gauge invariant variables read
\beq
\label{scalarbulkeqs}
\left.
\begin{aligned}
& \Box_6 \, \pi = 0 \\[3mm]
& \boxf \, h^{gi}_{ij} + 3 \, \d_{ij} \, \boxf \pi + 4 \, \de_{i} \de_{j} \, \pi = 0
\end{aligned}
\quad \right\} \qquad (\textrm{bulk})
\eeq
while the pure cod-1 junction conditions read
\beq
\label{scalarpurecod1xixi}
2 \, M_6^4 \, \bigg( 2 \, \de_{\bar{\mathbf{n}}}\evbhsi \, \pi - \boxf \, \hdvfn \bigg) + \frac{3}{2} \, M_5^3 \, \boxf \, \hp = 0 \qquad (\textrm{pure cod-1 brane}, \,\, \xi\xi)
\eeq
\vspace{2mm}
\beq
\label{scalarpurecod1ximu}
2 \, M_6^4 \, \bigg( \half \, \bn^i \bv^{j} \, \h{h}^{gi}_{ij} + \hdvfn^{\p} \, \bigg) - \frac{3}{2} \, M_5^3 \, \, \hp^{\p} = 0 \qquad (\textrm{pure cod-1 brane}, \,\, \xi\m)
\eeq
\vspace{3mm}
\beq
\label{scalarpurecod1dermunu}
2 M_6^4 \, \boxf \, \hdvfn + M_5^3 \bigg( \! - \frac{1}{2} \, \bv^{i} \bv^{j} \, \boxf \, \h{h}^{gi}_{ij} - \boxf \, \hp \bigg) = 0 \qquad (\textrm{pure cod-1 brane}, \,\, \demden)
\eeq
\begin{multline}
\label{scalarpurecod1tracemunu}
2 M_6^4 \, \bigg( \frac{3}{2} \, \de_{\bar{\mathbf{n}}}\evbhsi \, \pi + \frac{1}{2} \, \bv^{i} \bv^{j} \, \de_{\bar{\mathbf{n}}}\evbhsi \, h^{gi}_{ij} - \bn^i \bv^{j} \, \de_{\bar{\mathbf{v}}}\evbhsi \, h^{gi}_{ij} - \boxfi \, \hdvfn \bigg) + \\[1mm]
+ M_5^3 \, \bigg( \frac{1}{2} \, \bv^{i} \bv^{j} \, \, \boxf \h{h}^{gi}_{ij} + \frac{3}{2} \, \hp^{\p \p} + \boxf \, \hp \bigg) = 0 \qquad (\textrm{pure cod-1 brane}, \,\, \e_{\m\n})
\end{multline}
Moreover, the cod-2 junction conditions read
\beq
\left.
\begin{aligned}
\label{scalarcod2eqs}
\Bigg[ 4 \, M_6^4 \, \hdvfn^{\p} + M_6^4 \, \sin \bigg( \frac{\bla}{2 \Msf} \bigg) \Big( \h{h}^{gi}_{zz} - \h{h}^{gi}_{yy} \Big) - 3 \, M_5^3 \, \hp^{\p} \Bigg]_{0^+} \!= 0\,\,& \\[3mm]
2 \, \Mft \, \tan \bigg( \frac{\bla}{4 \Msf} \bigg) \, \boxf \, \hdvfn\Big\rvert_{0^+} = - \third \, \mcal{T}^{(4)} + \Mfs \, \boxf \, \pi^{(4)} &
\end{aligned}
\quad \right\} \qquad (\textrm{cod-2 brane})
\eeq
where the evaluation in $0^+$ is a shorthand for the evaluation on the side of the thin cod-1 brane $\hxi = 0^+$, and we expressed $\boxf \, \mcal{T}^{(4)}_{de}$ in terms of $\mcalT^{(4)}$ using the relation (\ref{cod2enmomtrace}). Note furthermore that the second of the cod-2 junction conditions can be equivalently expressed in terms of the bending of the cod-2 brane as follows
\beq
2 \, \Mft \, \sin \bigg( \frac{\bla}{4 \Msf} \bigg) \, \boxf \, \d\!\b^{(4)} = - \third \, \mcal{T}^{(4)} + \Mfs \, \boxf \, \pi^{(4)}
\eeq
since $\d\!\b^{(4)}$ and $\hdvfn\big\rvert_{0^+}$ are linked by the relation $\hdvfn\big\rvert_{0^+} = \cos(\bla/4 \Msf) \, \d\!\b^{(4)}$.

As we discuss in \cite{Sbisa':2014uza}, only two of the pure cod-1 junction conditions (\ref{scalarpurecod1xixi})--(\ref{scalarpurecod1tracemunu}) are independent if we take into account the second of the bulk equations (\ref{scalarbulkeqs}). Expressing the equations in terms of the master 
variables $\pi$ and $\hdvfn$, we obtain the following coupled system of differential equations
\begin{align}
\boxs \, \pi &= 0 \label{pibulkeq} \\[4mm]
2 \, M_6^4 \, \de_{\bar{\textbf{n}}}\evbhsi \, \pi + M_5^3 \, \boxfi \, \hp &= 0 \label{pipurecod1jc} \\
3 \, M_5^3 \, \hp^{\p}\Big\rvert_{0^+} &= \Bigg[ 4 \, M_6^4 \, \hdvfn^{\p} - 4 \, M_6^4 \, \sin \bigg( \frac{\bla}{2 \Msf} \bigg) \frac{\big( \de_z^2 - \de_y^2 \big)}{\boxf} \, \pi\evbbhsi  \Bigg]_{0^+} \label{picod2jc}
\end{align}
and
\begin{align}
\boxfi \, \hdvfn &= \half \, \de_{\bar{\textbf{n}}}\evbhsi \, \pi + 2 \, \de^2_{\hxi} \, \bigg( \frac{\de_{\bar{\textbf{n}}}}{\boxf}\evbhsi \, \pi \bigg) \label{normalbendingwaveeq} \\[2mm]
2 \, \Mft \, \tan \bigg( \frac{\bla}{4 \Msf} \bigg) \, \boxf \, \hdvfn\Big\rvert_{0^+} &= - \third \, \mcal{T}^{(4)} + \Mfs \, \boxf \, \pi^{(4)} \label{normalbendingjc}
\end{align}
Note that the cod-2 junction conditions (\ref{scalarcod2eqs}) act as boundary conditions at the thin cod-2 brane for the pure cod-1 junction conditions: (\ref{picod2jc}) acts as a boundary condition of the Neumann type for $\hp$, while (\ref{normalbendingjc}) acts as boundary condition of the Dirichlet type for $\boxf \, \hdvfn$.

The thin limit equations (\ref{pibulkeq})--(\ref{normalbendingjc}) were derived assuming that the bulk metric converges uniformly to a smooth limit configuration, while the embedding functions converge to a cuspy configuration. We showed in \cite{Sbisa':2014uza} that, in the $\Mfs = 0$ case, the thin limit equations are consistent for every form of source configuration; it is however easy to see that the same analysis holds also when $\Mfs \neq 0$. This result, together with the fact that the internal structure of the cod-2 brane do not appear in the thin limit equations, implies that the thin limit of the ribbon cod-2 brane inside the (already) thin cod-1 brane is well-defined in the nested branes realization of the Cascading DGP model (at least when considering first order perturbations around pure tension solutions). Moreover, since the embedding functions are continuous even in the thin limit and the bulk metric is smooth, the gravitational field on the cod-2 brane is finite for every form of the matter energy-momentum tensor on the cod-2 brane. This confirms that gravity in the Cascading DGP model is regularized by the cod-1 brane with induced gravity, as anticipated by \cite{deRham:2007xp,deRham:2007rw}.

\section{The critical tension and ghost-free regions in parameters space}
\label{The critical tension and ghost-free regions in parameters space}

We now focus on the dynamic of the metric master variable $\pi$ on the thin codimension-2 brane, where the critical tension is expected to emerge.

\subsection{The critical tension and its geometric interpretation}

\subsubsection{The critical tension}

Note that, if the background tension $\bla$ is non-vanishing, the cod-2 junction condition (\ref{normalbendingjc}) links the value of the ($\boxf$ of the) normal component of the bending $\hdvfn$ on the side of the cod-2 brane with the value of the $\pi$ field on the cod-2 brane, and with the trace of the matter energy-momentum tensor on the cod-2 brane. On the other hand, the $\xi\xi$ component of the pure cod-1 junction conditions (equation (\ref{scalarpurecod1xixi})) links the ($\boxf$ of the) normal component of the bending $\hdvfn$ to the $\pi$ field on the cod-1 brane, and to the derivative of $\pi$ normally to the cod-1 brane. Evaluating the latter equation on the side of the cod-2 brane (\emph{i.e.} considering the $\hxi \rightarrow 0^+$ limit of the equation (\ref{scalarpurecod1xixi})), we obtain by continuity a relationship between the value of $\boxf \hdvfn$ on the side of the cod-2 brane, the value of $\boxf \pi$ on the cod-2 brane and the derivative of $\pi$ normally to the cod-1 brane on the side of the cod-2 brane
\beq
\label{freedomfreedom}
4 \, M_6^4 \, \de_{\bar{\mathbf{n}}} \, \pi\Big\rvert_{0^+} - 2 \, M_6^4 \, \boxf \, \hdvfn\Big\rvert_{0^+} + \frac{3}{2} \, M_5^3 \, \boxf \, \hp\Big\rvert_{0^+} = 0
\eeq
where the latter equation contains function of the 4D variables $\chd$ only, and we introduced the notation
\beq
\de_{\bar{\mathbf{n}}} \, \pi\Big\rvert_{0^+} = \de_{\bar{\mathbf{n}}} \, \pi\Big\rvert_{\bvfd(\hxi = 0^+)}
\eeq
Therefore, we can then use the two equations (\ref{normalbendingjc}) and (\ref{freedomfreedom}) to obtain a master equation for the field $\pi$, using the fact that by continuity of the $\pi$ field we have $\boxf \, \pi^{(4)} = \boxf \, \hp\big\rvert_{0^+}$. In fact, expressing $\boxf \, \hdvfn\big\rvert_{0^+}$ in terms of $\pi^{(4)}$ and $\mcal{T}^{(4)}$ using the equation (\ref{normalbendingjc}), and inserting the resulting relation in the equation (\ref{freedomfreedom}), we get
\beq
\label{Brigitte}
4 \, M_6^4 \, \tan \bigg( \frac{\bla}{4 M_6^4} \bigg) \, \de_{\bar{\mathbf{n}}} \, \pi\Big\rvert_{0^+} + \bigg[ \frac{3}{2} \, M_5^3 \, \tan \bigg( \frac{\bla}{4 M_6^4} \bigg) - m_6 \, M_4^2 \bigg] \, \boxf \, \hp\Big\rvert_{0^+} = - \frac{m_6}{3} \, \mcal{T}^{(4)}
\eeq
This equation is exact (at first order in perturbations), since we didn't take any ``decoupling limit'' to obtain it; considering the thin limit on the cod-2 brane allowed us to find a master equation for the field $\pi$ and its derivatives at the cod-2 brane.

Despite the equation (\ref{Brigitte}) involves only the value of the field $\pi$ near the cod-2 brane, the presence of the normal derivative $\de_{\bar{\mathbf{n}}} \, \pi$ implies that to find a solution of (\ref{Brigitte}) we have to solve the bulk equations and the cod-1 junction conditions, or in other words we still need to solve the complete system of differential equations for $\pi$ and $\hdvfn$. However, it is possible to look for an approximate description which ``decouples'' the dynamics on the cod-2 brane from the dynamics in the bulk and on the cod-1 brane, with the hope to find a master equation which describes the behaviour of $\pi$ on the cod-2 brane. We consider in fact the following ``4D limit''
\beq
\label{4Dlimit}
\begin{aligned}
\hspace{2mm} \lvert m_6 \, \de_{\bar{\mathbf{n}}} \rvert &\ll \lvert \boxf \rvert \\[2mm]
\hspace{2mm} \lvert m_5 \, \de_{\bar{\mathbf{n}}} \rvert &\ll \lvert \boxf \rvert
\end{aligned}
\eeq
which implies that, in the left hand side of equation (\ref{Brigitte}), we can neglect the first term compared to the second term and to the third term: we then obtain
\beq
\label{critical tension equation}
3 M_4^2 \, \bigg[ \, 1 - \frac{3}{2} \, \frac{m_5}{m_6} \, \tan \bigg( \frac{\bla}{4 M_6^4} \bigg) \bigg] \,\, \boxf \, \pi^{(4)} = \mcal{T}^{(4)}
\eeq
This is the effective master equation which describes the behaviour of $\pi$ field on the cod-2 brane in the 4D limit (note that this limit is different from the decoupling limit considered in \cite{deRham:2007xp,deRham:2007rw} which corresponds to an effective 5D description). Crucially, in this equation the numerical coefficient which multiplies $\boxf \, \pi^{(4)}$ changes sign when $\bla$ becomes equal to the \emph{critical tension}
\beq
\label{critical tension}
\bla_c = 4 M_6^4 \, \arctan \bigg( \frac{2}{3} \, \frac{m_6}{m_5} \bigg)
\eeq

\subsubsection{Ef\mbox{}fective action and ghosts}

The sign of the coefficient multiplying $\boxf \, \pi^{(4)}$ in equation (\ref{critical tension equation}) is closely related to the fact that the field $\pi$ is an effective ghost or not. In fact, the equation (\ref{critical tension equation}) tells us that, in the 4D limit, the dynamics of the field $\pi^{(4)}$ is described by an effective 4D action which is proportional to
\beq
\label{Alessandra}
S^{(2)}_{\pi^{(4)}} = \int \! d^4 \ch \,\, \bigg[ \, \frac{K}{2} \, \de_{\m} \pi^{(4)} \de^{\m} \pi^{(4)} + \pi^{(4)} \mcal{T}^{(4)} \bigg]
\eeq
where we indicated
\beq
\label{kineticterm}
K \equiv 3 M_4^2 \, \bigg[ \, 1 - \frac{3}{2} \, \frac{m_5}{m_6} \, \tan \bigg( \frac{\bla}{4 M_6^4} \bigg) \bigg]
\eeq
The effective 4D action is in general obtained by integrating out of the (quadratic approximation of the) general action (\ref{CascadingDGP6D}) all the other fields using the bulk equations and the junction conditions, and imposing the conditions (\ref{4Dlimit}). However, in practice we just need to determine the value of the proportionality constant between the true effective action and (\ref{Alessandra}), which can be recognized from the coupling of $\pi^{(4)}$ with the matter. Expanding at quadratic order around the Minkowski spacetime the term in the general action which expresses the gravity-matter coupling, we get
\beq
\int \! d^4 \ch \, \sqrt{- g^{(4)}} \, \mscr{L}_m \simeq \int \! d^4 \ch \, h^{(4)}_{\m\n} \, \mcal{T}_{(4)}^{\m\n} = \int \! d^4 \ch \, \Big( \pi^{(4)} \mcal{T}^{(4)} + \mscr{H}^{(4)}_{\m\n} \mscr{T}_{(4)}^{\m\n} \Big)
\eeq
whose scalar part indicates that the action (\ref{Alessandra}) is indeed the correct 4D effective action for $\pi^{(4)}$. Since, with our choice of the metric signature, a field which obeys an action of the form (\ref{Alessandra}) is a ghost if $K > 0$ while it is a healthy field if $K < 0$, we conclude that the field $\pi^{(4)}$ in the nested branes realization of the 6D Cascading DGP model is an effective 4D ghost if the background tension is smaller than the critical tension $\blac$, while it is a healthy effective 4D field if the background tension is bigger than $\blac$. From the point of view of the action, we can say that, integrating out the other fields in the scalar sector and imposing the decoupling limit, we generate a $\bla$-dependent contribution
\beq
S^{(2)}_{\bla} = \int \! d^4 \ch \,\, \bigg[ \, - \frac{9}{4} \, M_4^2 \, \frac{m_5}{m_6} \, \tan \bigg( \frac{\bla}{4 M_6^4} \bigg) \, \de_{\m} \pi^{(4)} \de^{\m} \pi^{(4)} \, \bigg]
\eeq
to the kinetic term of $\pi^{(4)}$ which is added to the 4D part of the general Lagrangian 
\beq
S_{4}^{(2)} = \int \! d^4 \ch \,\, \bigg[ \, \frac{3 M_4^2}{2} \, \de_{\m} \pi^{(4)} \de^{\m} \pi^{(4)} + \pi^{(4)} \mcal{T}^{(4)} \, \bigg]
\eeq
and cures the ghost if $\bla > \blac$.

Note that, strictly speaking, to claim that the field $\pi^{(4)}$ is a ghost we should perform a Hamiltonian analysis, since from a Hamiltonian point of view the system we are studying is a constrained system. To see why this may be relevant, consider the limit $\bla \rightarrow 0$ of the action (\ref{Alessandra}). Performing this limit we obtain the action for the scalar sector of (4D) GR: if the above reasoning regarding the sign of the kinetic term were conclusive, we should conclude that GR itself has a ghost (this is known as the conformal factor problem in GR). However, a careful Hamiltonian analysis of GR permits to show that the constrained structure of the theory renders the $\pi^{(4)}$ field non-propagating, so GR is ghost-free despite the wrong sign of the kinetic term of $\pi^{(4)}$ \cite{Schleich:1987,Mazur:1990,Berkhahn:2012wg,Hassan:2010ys,Hassan:2011zd}. The analysis of the GR action does not extend to the 6D Cascading DGP, since in the former case $\pi^{(4)}$ is the trace part of a 4D graviton while in the latter case it is the 4D trace of a 6D graviton, and the Hamiltonian analysis should be different in the two cases. Nevertheless, we feel that the result of $\pi^{(4)}$ being a ghost when $\bla < \blac$ and healthy when $\bla > \blac$ should be confirmed by a full-fledged Hamiltonian analysis. We leave this for future work. From another point of view, it is important to remember that we obtained this result at first order in perturbations, so the presence/absence of the ghost should be confirmed at full non-linear level.

\subsubsection{Geometrical interpretation of the critical tension}

We can now understand geometrically what is the role of the background tension concerning the dynamics of $\pi^{(4)}$ and the sign of its kinetic term, and in particular why a critical tension emerges at all. First of all, note that the 4D limit equation (\ref{critical tension equation}) for the $\pi^{(4)}$ field can be obtained directly from the the equations (\ref{normalbendingjc}) and (\ref{freedomfreedom}) if we neglect the term $\Msf \, \de_{\bar{\mathbf{n}}} \pi$ in (\ref{freedomfreedom}), so we can consider the following system of equations
\begin{align}
\label{normalbendingjcdeclimit}
2 \, M_5^3 \, \tan \bigg( \frac{\bla}{4 M_6^4} \bigg) \, \, \boxf \, \hdvfn\Big\rvert_{0^+} &= M_4^2 \, \boxf \, \pi^{(4)} - \frac{1}{3} \, \mcal{T}^{(4)} \\[2mm]
\label{freedomfreedomdeclimit}
2 \, M_6^4 \, \boxf \, \hdvfn\Big\rvert_{0^+} &= \frac{3}{2} \, M_5^3 \, \boxf \, \hp\Big\rvert_{0^+}
\end{align}
as the 4D limit of the system (\ref{normalbendingjc})-(\ref{freedomfreedom}). Furthermore, it is convenient to express these equations in terms of objects which have a clear geometrical meaning also in the thin limit, and in particular it is useful to write the equation \eqref{normalbendingjc} in terms of the bending mode $\d\!\b^{(4)}$ of the cod-2 brane in the bulk. The equations (\ref{normalbendingjcdeclimit})-(\ref{freedomfreedomdeclimit}) then read
\begin{align}
\label{thinkdeclimitgeom}
6 \, M_5^3 \, \sin \bigg( \frac{\bla}{4 M_6^4} \bigg) \, \, \boxf \, \d\!\b^{(4)} - 3 M_4^2 \, \boxf \, \pi^{(4)} &= - \mcal{T}^{(4)} \\[2mm]
\label{freedomfreedomdeclimitgeom}
2 \, M_6^4 \, \boxf \, \hdvfn\Big\rvert_{0^+} &= \frac{3}{2} \, M_5^3 \, \boxf \, \hp\Big\rvert_{0^+}
\end{align}
and need to be completed with the continuity conditions
\begin{align}
\label{bendinglink}
\hdvfn\Big\rvert_{0^+} &= \cos \bigg( \frac{\bla}{4 \Msf} \bigg) \, \d\!\b^{(4)} \\[2mm]
\label{pilink}
\boxf \, \hp\Big\rvert_{0^+} &= \boxf \, \pi^{(4)}
\end{align}
The equation (\ref{bendinglink}) expresses the fact that, since the components of the embedding function $\dvf^A$ are continuous (the cod-1 brane ``does not break''), the movement of the cod-2 brane and the movement of the cod-1 brane near the cod-2 brane are linked. However, since $\hdvfn\big\rvert_{0^+}$ is constructed from the cod-1 embedding by projecting on the normal vector, and the background normal vector depends on the background tension, $\hdvfn\big\rvert_{0^+}$ and $\d\!\b^{(4)}$ are linked in a $\bla$-dependent way.

We can interpret the system of equations (\ref{thinkdeclimitgeom})-(\ref{pilink}) in the following way. The equation (\ref{thinkdeclimitgeom}) tells us that the presence of matter on the cod-2 brane (represented by $\mcalT^{(4)}$) has two effects: on one hand, it excites the metric perturbations on the cod-2 brane (represented by $\pi^{(4)}$) via the 4D induced gravity term, and in a ghostly way. On the other hand, since the 4D brane is actually part of a 6D set-up and in fact embedded into a 5D cod-1 brane, $\mcalT^{(4)}$ excites also the movement of the cod-2 brane in the bulk (represented by $\d\!\b^{(4)}$), this time in a healthy way. However, it does so in a $\bla$-dependent way, and this excitation mechanism is the more efficient the larger the background tension, while it is completely inefficient when $\bla$ is very small. As we already mentioned, the equation (\ref{bendinglink}) instead tells us that, since the cod-2 brane is embedded inside the cod-1 brane, the movement of the cod-2 brane ``drags'' the cod-1 brane as well; therefore the matter on the cod-2 brane indirectly excites $\hdvfn\big\rvert_{0^+}$. Passing from $\d\!\b^{(4)}$ to $\hdvfn\big\rvert_{0^+}$ we gain an additional $\bla$-dependence, but the sign does not change and so $\mcalT^{(4)}$ excites $\hdvfn\big\rvert_{0^+}$ in a healthy way. In turn, considering now the equation (\ref{freedomfreedomdeclimitgeom}), $\hdvfn\big\rvert_{0^+}$ excites the metric perturbations (expressed by the field $\hp$) on the cod-1 brane via the 5D induced gravity term, still in a healthy way. By continuity of the $\hp$ field (equation (\ref{pilink})), the perturbation of $\hp$ finally induces the perturbation of $\pi^{(4)}$, in a healthy way.

To sum up, the presence of matter on the cod-2 brane excites the field $\pi^{(4)}$ via two separate channels: it does so directly, because of the 4D induced gravity term, and indirectly via the bending of the cod-1 brane, because of the 5D induced gravity term. Furthermore, we saw above that the first channel excites $\pi^{(4)}$ in a ghostly and $\bla$-independent way, while the second channel excites $\pi^{(4)}$ in a healthy and $\bla$-dependent way. The fact that the field $\pi^{(4)}$ is a ghost or not is decided by the fact that the first or the second channel is more efficient than the other. In particular, the existence of the critical tension is due to the competition between these two channels, and its value corresponds to the tension where the two channels are equally efficient. Note finally that the existence of the second channel is entirely due to the higher dimensional structure of the theory. This is seen from the point of view of the action as the fact that the healthy part of the effective 4D kinetic term (which cures the presence of the ghost for $\bla > \blac$) is created by integrating out the other fields in the 6D and 5D parts of the total action.

\subsection{Ghost-free regions in parameters space}
\label{Ghost-free regions in parameters space}

The results of the previous section seem to be at odds with the findings of \cite{deRham:2010rw} (which agree with \cite{deRham:2007xp}), where the following kinetic part of the 4D effective action for the field $\pi^{(4)}$ was found 
\beq
S^{\textup{dRKT}}_{\textup{cod-2}} = \int \! d^4 \chi \, \frac{3 M_4^2}{4} \bigg( \frac{3 \bla}{2 m_6^2 M_4^2} - 1 \bigg) \, \pi^{(4)} \boxf \pi^{(4)}
\eeq
which has exactly the same structure of (\ref{Alessandra})--(\ref{kineticterm}) although with a different value for the critical tension. To compare the two results, note first of all that our conventions differ slightly from those of \cite{deRham:2010rw}, since in our cod-1 junction conditions (\ref{junctionconditionseq}) the mass $\Msf$ is multiplied by 2, while it is not so in their case. Rescaling $\Msf \to 2 \Msf$ in their result, to normalize the conventions, the critical tension in their case reads
\beq
\label{critical tension theirs}
\blac^{\textup{dRKT}} = \frac{8 \, m_6^2 \, M_4^2}{3}
\eeq
which is different from the value (\ref{critical tension}) we find in our analysis. Referring to the discussion in section \ref{The nested branes realization} and in \cite{Sbisa':2014uza}, one may suggest that the two results differ because we are considering different realizations of the Cascading DGP (i.e.~different choices for the internal structures), so that in truth we are considering different models. However, as we show in appendix \ref{The brane-based approach}, there is a coordinate transformation which links our analysis to that of \cite{deRham:2010rw}, and so we are indeed studying the same set-up although in different coordinate systems (we use a bulk-based approach, while \cite{deRham:2010rw} uses a brane-based approach). Note that, strictly speaking, there is no contradiction with the estimate of \cite{deRham:2007xp}, since that was found in the 5D decoupling limit which applies only for small values of the ratio $m_6/m_5$. Instead, the result is in sharp contradiction with \cite{deRham:2010rw}.

This difference has important consequences for the phenomenological viability of the model. Note that, when $m_6 \ll m_5$ (i.e.~$\Msf \ll M_{5}^{6}/\Mfs$, for example when the background tension is very small $\bla \ll 1$), our result reads
\beq
\bla_c \simeq \frac{8 \, m_6 \, M_6^4}{3 \, m_5} = \frac{8 \, m_6^2 \, M_4^2}{3} = \blac^{\textup{dRKT}}
\eeq
and so coincides with the result of \cite{deRham:2010rw}. However, when $m_6$ is not negligible with respect to $m_5$ the two results are different, and become dramatically so when $m_6 \gg m_5$. It is illuminating to consider the ratio between the critical tension and the maximum tension $\la_M = 2 \pi M_{6}^{4}$ which can be placed on the cod-2 brane: in our case we get
\beq
\frac{\blac}{\bla_M} = \frac{2}{\pi} \, \arctan \bigg( \frac{2}{3} \, \frac{m_6}{m_5} \bigg)
\eeq
while with the result of \cite{deRham:2010rw} we get
\beq
\frac{\blac^{\textup{dRKT}}}{\bla_M} = \frac{4}{3 \pi} \, \frac{m_6}{m_5}
\eeq
It is apparent that with our result $\blac$ remains smaller than $\bla_M$ for every value of the parameters $m_5$ and $m_6$, while with the result of \cite{deRham:2010rw} this is true only when $m_6 \lesssim m_5$. Crucially, our result implies that, for every value of the free parameters of the model, there is an interval of values for the background tension such that $\pi^{(4)}$ is not a ghost, and so the model is phenomenologically viable. The result of \cite{deRham:2010rw}, on the other hand, implies that the $m_6 > m_5$ region in parameters space is phenomenologically ruled out. Since for $m_6 \ll m_5$ the gravitational field Cascades from 6D to 5D to 4D progressing from very large to very small scales, while in the $m_6 \gg m_5$ there is a direct transition from 6D to 4D, it was concluded in \cite{deRham:2007xp,deRham:2010rw} that the latter behaviour of the gravitational field is ruled out. Our result instead implies that both behaviours are viable.

\subsubsection{Thin limit and pillbox integration}

It is therefore very important to understand which of the two results is the correct one, or at least understand why two different results are obtained. As we mentioned above, we derived the critical tension using the $\xi\xi$ component of the cod-1 junction conditions (\ref{scalarpurecod1xixi}) and the cod-2 junction condition (\ref{normalbendingjc}), where the latter comes from the pillbox integration across the ribbon brane of the (4D trace of the) derivative part of the cod-1 induced gravity term (see \cite{Sbisa':2014uza} for the detailed derivations)
\beq
\label{pillboxfantasy}
\lim_{l_2 \to 0^+} \int_{-l_2}^{+l_2} \! d \hxi \, \Mft \, \bigg[ \!- \frac{1}{2} \, \bv^{i} \bv^{j} \, \boxf \, \h{h}^{gi}_{ij} - \boxf \, \hp + \bv_{i}^{\p} \, \boxf \, \hdvf^{i}_{gi} \bigg] = M_4^2 \, \boxf \, \pi^{(4)} - \frac{1}{3} \, \mcal{T}^{(4)}
\eeq
Of these two equations, the cod-2 junction condition is the most delicate to derive, since the pillbox integration in the nested branes with induced gravity set-ups is very subtle \cite{Sbisa':2014uza}: it is quite natural to investigate first if the difference can be traced back to the way the pillbox integration is performed. In this respect, note that the result of \cite{deRham:2007xp,deRham:2010rw} were reproduced in our approach if we performed the pillbox integration (\ref{pillboxfantasy}) in such a way to obtain the following change in the equation (\ref{normalbendingjc})
\beq
\label{transition}
2 \, M_5^3 \, \tan \bigg( \frac{\bla}{4 M_6^4} \bigg) \, \, \boxf \, \hdvfn\Big\rvert_{0^+} \qquad \to \qquad \frac{\bla}{2 m_6} \, \boxf \, \hdvfn\Big\rvert_{0^+}
\eeq

To perform the pillbox integration in a rigorous way, in \cite{Sbisa':2014uza} we introduced a sequence of brane configurations indexed by $n \in \mathbb{N}$ where the width $l_2^{_{[n]}}$ of the ribbon brane converges to zero $l_2^{_{[n]}} \to 0^+$, while the pillbox integration of the source configurations $\h{T}_{ab}^{_{[n]}}$ is independent of $n$ (which in particular means that $f_{1}^{_{[n]}}$ and $f_{2}^{_{[n]}}$ are two representations of the Dirac delta, and $\mcalT_{ab}^{_{[n]}}$ converges to $\d(\hxi) \, \d_{a}^{\,\,\, \m} \d_{b}^{\,\,\, \n} \mcalT_{\m\n}^{_{(4)}}$). For each value of $n$, we associated to the source configuration the corresponding solution of the equations of motion for the bending and for the metric perturbations, so to the sequence of source configurations we associated a sequence of geometric configurations $h^{_{[n]}}_{AB}$, $\bvf^A_{^{[n]}}$ and $\dvf^A_{^{[n]}}$. The pillbox integration was therefore defined as the limit
\beq
\lim_{n \to \infty} \int_{-l_{2}^{[n]}}^{+l_{2}^{[n]}} \! d \hxi
\eeq
of the relevant equations (the $\m\n$ components of the cod-1 junction conditions). We do exactly the same here, with the only difference that now we indicate the limit configurations without the $\infty$ subscript/superscript (which characterize the limit configurations in \cite{Sbisa':2014uza}), so for example
\begin{align}
\bvf^i_{[n]} &\xrightarrow[n \to \infty]{} \bvf^i & \dvf^i_{[n]} &\xrightarrow[n \to \infty]{} \dvf^i & \pi_{[n]} &\xrightarrow[n \to \infty]{} \pi
\end{align}
It is clear that the only terms which give a non-negligible contribution to the pillbox integration are those which diverge on the cod-2 brane in the thin limit. Crucially, it is not possible to say a priori which terms will diverge, but the only thing we can do is to propose an ansatz for the behaviour of the perturbation fields in the thin limit, perform the pillbox integration and a posteriori verify that the ansatz is consistent.

In \cite{Sbisa':2014uza}, we proposed an ansatz in which the bulk metric converges to a smooth configuration, while the embedding converges to a cuspy configuration. This is consistent with the properties of the pure tension (background) solutions, which display exactly this behaviour, and is linked to the idea that the cusp of the embedding functions supports all the singularity in the geometric configuration. This is the same ansatz we used in the present paper to derive the thin limit equations of section \ref{Thin limit equations of motion}. We showed in \cite{Sbisa':2014uza} that this ansatz produces a consistent system of thin limit equations, so it is consistent itself. According to this ansatz, the only terms which give a non-vanishing contribution to the pillbox integration are those which contain fields derived twice with respect of $\hxi$, either embedding functions (background or bending modes) or bulk metric perturbations evaluated on the cod-1 brane. On the other hand, derivatives of every order of the bulk perturbations with respect to the bulk coordinates are smooth, and so remain bounded when evaluated on the cod-1 brane. It follows that the only term in (\ref{pillboxfantasy}) which contributes is $\bv_{i}^{\p} \, \boxf \, \hdvf^{i}_{gi}$, since $\bv_{i}^{\p} = \bvf_{i}^{\p\p}$, so the cod-2 junction condition (\ref{normalbendingjc}) reads
\beq
\label{Palinka}
\lim_{n \rightarrow + \infty} M_5^3 \int_{-l_{2}^{[n]}}^{+l_{2}^{[n]}} \! d \hxi \,\, \bvf_{i}^{[n] \, \p\p} \, \boxf \, \hdvf^{i\,[n]}_{gi} = M_4^2 \, \boxf \, \pi^{(4)} - \frac{1}{3} \, \mcal{T}^{(4)}
\eeq

\subsubsection{Pillbox integration and convergence properties}
\label{Pillbox integration and convergence properties}

The subtle point is the evaluation of the integral in the left hand side of the previous equation\footnote{We take the $\boxf$ out of the integral, since the functions are smooth in the 4D directions also in the thin limit.}
\beq
\label{Vittoria}
\mscr{I} = \lim_{n \rightarrow + \infty} \int_{-l_{2}^{[n]}}^{+l_{2}^{[n]}} \! d \hxi \,\, \bvf_{i}^{[n] \, \p\p} \, \hdvf^{i\,[n]}_{gi}
\eeq
In \cite{Sbisa':2014uza}, we used the relation
\beq
\bvf_{i}^{[n] \, \p\p} \,  \hdvf^{i\,[n]}_{gi} = \Big( \bvf_{i}^{[n] \, \p} \,  \hdvf^{i\,[n]}_{gi} \Big)^{\p} - \bvf_{i}^{[n] \, \p} \,  \hdvf^{i \,[n] \, \p}_{gi}
\eeq
and the fact that the second term on the right hand side does not diverge in the thin limit, since it does not contain second derivatives, so its pillbox integration tends to zero when $n \rightarrow + \infty$. The integral of the first term in the right hand side is trivial, and we obtain
\beq
\mscr{I} = 2 \lim_{n \rightarrow + \infty} \Big[ \bvf_{i\,[n]}^{\p} \,  \hdvf^{i\,[n]}_{gi} \Big]_{l_2^{[n]}} = 2 \lim_{n \rightarrow + \infty} \Big[ \Zp_{[n]} \,  \hdvf^{z\,[n]}_{gi} \Big]_{l_2^{[n]}} = 2 \, \sin \bigg( \frac{\bla}{4 M_6^4} \bigg) \, \d\!\b^{(4)}
\eeq
where we used the fact that $\hdvf^{y}_{gi}$ vanishes in $\hxi = 0$ since it is odd and continuous. We can express this result in terms of the normal bending $\hdvfn$, since $\hdvfn\big\rvert_{0^+} = \Yp\big\rvert_{0^+} \, \d\!\b^{(4)}$, to obtain
\beq
\label{RiminiRimini1}
\mscr{I} = 2 \, \tan \bigg( \frac{\bla}{4 M_6^4} \bigg) \, \hdvfn \Big\rvert_{0^+}
\eeq
We refer to this evaluation of the integral $\mscrI$ as ``route A''. However, we may take a different route (which we call ``route B''): in fact, using the relations (\ref{Patty}), (\ref{Jerry}) and (\ref{Irioth}) we deduce that
\beq
\bvf_{i}^{[n] \, \p\p} = S^{\p}_{[n]} \, \bn_{i}^{[n]}
\eeq
and therefore we can write the integrand of (\ref{Vittoria}) in terms of the normal component of the bending 
\beq
\bvf_{i}^{[n] \, \p\p} \, \hdvf^{i\,[n]}_{gi} = S^{\p}_{[n]} \, \hdvfn^{[n]}
\eeq
Since, by equation (\ref{satisfaction}), $S^{\p}_{[n]}$ is proportional to a realization of the Dirac delta
\beq
S^{\p}_{[n]} = \frac{\bla}{2 M_6^4} \, f_{1}^{[n]}(\hxi)
\eeq
we may be tempted to use the defining property of the Dirac delta
\beq
\label{classicdeltarep}
\lim_{n \rightarrow + \infty} \, \int_{-\infty}^{+\infty} \! d \hxi \,\, f_{1}^{[n]}(\hxi) \, \mcal{F}(\hxi) = \mcal{F}(0)
\eeq
to evaluate the integral $\mscr{I}$ as follows
\beq
\label{RiminiRimini2}
\mscr{I} = \frac{\bla}{2 M_6^4} \, \lim_{n \rightarrow + \infty} \int_{-l_{2}^{[n]}}^{+l_{2}^{[n]}} \! d \hxi \,\, f_{1}^{[n]}(\hxi) \,\, \hdvfn^{[n]} = \frac{\bla}{2 M_6^4} \, \hdvfn\Big\rvert_{0^+}
\eeq
Comparing (\ref{RiminiRimini1}) to (\ref{RiminiRimini2}), it is evident that, using the route B to perform the integral $\mscrI$ instead of the route A, the term $\tan \big( \bla/4 M_6^4 \big)$ is substituted by $\bla/4 M_6^4$. It follows that, if we use the route B to derive the cod-2 junction condition (\ref{normalbendingjc}) instead of the route A, we indeed generate the change (\ref{transition}) in the equation (\ref{normalbendingjc}) and so we obtain exactly the result of \cite{deRham:2007xp,deRham:2010rw} for the value of the critical tension. 

However, it is possible to see that the route B is not mathematically well justified. In fact, its central point is the use of the property of the Dirac delta (\ref{classicdeltarep}) where the function $\mcal{F}(\hxi)$ is
\beq
\mcal{F}(\hxi) = \hdvfn^{[n]}
\eeq
The use of the formula (\ref{classicdeltarep}) with this identification involves a subtlety, since $\hdvfn^{[n]}$ is a sequence of functions: the usual proof of (\ref{classicdeltarep}) assumes that $\mcal{F}$ is a continuous function which is \emph{independent} of $n$. We proved in \cite{Sbisa':2014uza} that (\ref{classicdeltarep}) holds even when $\mcal{F}$ is replaced by a sequence $\mcal{F}_{n}$ of continuous functions, provided it converges uniformly to a continuous function $\mcal{F}_{\infty}$ (see \cite{Rudin} for the standard definitions of pointwise and uniform convergence). More precisely, in this case we have
\beq
\label{unifdeltarep}
\lim_{n \rightarrow + \infty} \, \int_{-\infty}^{+\infty} \! d \hxi \,\, f_{1}^{[n]}(\hxi) \, \mcal{F}_{[n]}(\hxi) = \mcal{F}_{\infty}(0)
\eeq
However, there is no guarantee that this relation holds if $\mcal{F}_{n}$ converges pointwise to some function (continuous or not). To understand why, we observe that the idea behind the formula (\ref{classicdeltarep}) is that, since $f_{1}^{_{[n]}}$ is peaked around $\hxi = 0$, it probes the function $\mcal{F}$ only around $\hxi = 0$. If $\mcal{F}$ is continuous, in the $n \rightarrow + \infty$ limit it can be considered nearly constant in the $\hxi$-interval where $f_{1}^{_{[n]}}$ is peaked, and so it can be taken out of the integral. The same happens for the formula (\ref{unifdeltarep}), since the behaviour of $\mcal{F}_{^{[n]}}$ near $\hxi = 0$ is under control if $\mcal{F}_{^{[n]}}$ converges uniformly to a continuous function. On the other hand, if the convergence of $\mcal{F}_{^{[n]}}$ is not uniform, $\mcal{F}_{^{[n]}}$ may develop a non-trivial behaviour (for example, a peak) around $\hxi = 0$ in the $n \rightarrow + \infty$ limit, as much as $f_{1}^{_{[n]}}$ does. In this case, by no means it can be considered constant and taken out of the integral, since its singular behaviour may contribute in a non-trivial way to the integral even in the limit.

\begin{figure}[htp!]
\begin{center}
\includegraphics{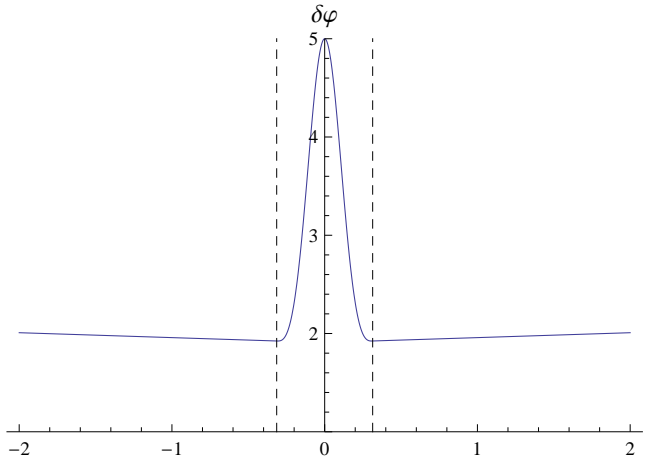}
\caption[The normal component of the bending]{The normal component of the bending $\hdvfn^{[n]}$ in the case of a pure tension perturbation. The vertical dashed lines are the boundaries of the ribbon brane.}
\label{normalbending}
\end{center}
\end{figure}

This is in fact what happens in our case. Remember that, by definition, $\hdvfn^{[n]}$ is the projection of the bending modes on the background normal vector
\beq
\hdvfn^{[n]} = \bn_{i}^{[n]} \, \hdvf^{i \, [n]}_{gi} = \Yp_{[n]} \, \hdvf^{z \, [n]}_{gi} - \Zp_{[n]} \, \hdvf^{y \, [n]}_{gi}
\eeq
The bending modes $\hdvf^{_{z \, [n]}}_{^{gi}}$ and $\hdvf^{_{y \, [n]}}_{^{gi}}$ necessarily have to converge to continuous functions ($\hdvf^{_{z}}_{^{gi}}$ and $\hdvf^{_{y}}_{^{gi}}$), otherwise the cod-1 brane would break into two pieces when the ribbon brane becomes thin. Actually, the functions $\hdvf^{_{y \, [n]}}_{^{gi}}$ and $\hdvf^{_{y}}_{^{gi}}$ vanish in $\hxi = 0$ since they are continuous and odd (as a consequence of the $\mathbb{Z}_2$ symmetry present inside the cod-1 brane), and so the normal component of the bending around the mathematical cod-2 brane reads
\beq
\hdvfn^{[n]} \simeq \Yp_{[n]} \, \hdvf^{z \, [n]}_{gi} \qquad \qquad \textrm{for} \qquad \qquad \hxi \simeq 0
\eeq
On the other hand, the background embedding $\Yp_{^{[n]}}$ displays a non-trivial behaviour inside the ribbon brane. This is easily seen in the numerical plot in figure \ref{background embedding Y} of appendix \ref{Thin limit of the background solutions}, obtained in the case of a pure tension perturbation (using the explicit realizations (\ref{fnexplicit}) and (\ref{epnexplicit}) for $n = 10$, and the same values of the free parameters as in section \ref{Choice of the free parameters}). The crucial point for the discussion above is that $\Yp_{^{[n]}}$ has a peak localized inside the ribbon brane. As a consequence of this, also the normal component of the bending displays the same peaked behaviour, which is manifest in the plot in figure \ref{normalbending} (obtained again for a pure tension perturbations with the same choices of figure \ref{background embedding Y}).
One may suggest that this peak becomes less and less important as $n$ gets bigger, and do not contribute in the $n \to \infty$ limit. We address this point in section \ref{Numerical check}. For the time being, we just show that the sequence of functions $\hdvfn^{[n]}$ cannot converge uniformly to its limit configuration, and so the route B to evaluate the pillbox integral is not mathematically justified. Note in fact that, as we show in appendix \ref{Thin limit of the background solutions}, the sequence $\Yp_{^{[n]}}$ converges to the \emph{discontinuous} limit configuration (\ref{felicita2}). This implies that also the sequence $\hdvfn^{_{[n]}}$ in general converges to a discontinuous configuration. However, by hypothesis the functions $\hdvfn^{_{[n]}}$ are smooth for every (finite) value of $n$. It is a general property that, if a sequence of smooth functions converges uniformly, then the limit configuration is (at least) continuous. This implies that the convergence of the sequence $\hdvfn^{_{[n]}}$ is not uniform but merely pointwise, and so the formula (\ref{RiminiRimini2}) is not justified.

\subsection{Numerical check}
\label{Numerical check}

In light of the discussion above, we propose that the reason for the difference between the result of the route A and of the route B lies in the fact that the route B does not take properly into account the singular structure of the fields at the cod-2 brane. To support this claim, we estimate numerically the pillbox integration
\beq
\label{Vittoria2}
\mscr{I} = \frac{\bla}{2 M_6^4} \, \lim_{n \rightarrow + \infty} \int_{-l_{2}^{[n]}}^{+l_{2}^{[n]}} \! d \hxi \,\, f_{1}^{[n]}(\hxi) \,\, \hdvfn^{[n]}
\eeq
in a particular case where the solution for the bending modes and for the metric perturbations is known exactly (also) \emph{inside} the thick cod-2 brane: the pure tension perturbation case. Since the solution is known exactly, we don't need to make any hypothesis on the behaviour and on the convergence properties of the perturbation fields: we can perform explicitly the integration in the right hand side of (\ref{Vittoria2}) for several values of $n$, and estimate the value of the limit $\mscr{I}$ by studying the asymptotic behaviour at large $n$. It is worthwhile to point out that the system of equations (\ref{pibulkeq})--(\ref{normalbendingjc}) (and therefore the equation (\ref{freedomfreedom})) were derived assuming $\dla = 0$, so the pure tension case is not directly related to the existence of the critical tension (a pure tension perturbation does not excite $\pi$). Nevertheless, concerning the integral $\mscrI$, the behaviour of the bending modes in the pure tension case is closely related to their behaviour in the general case $\mcalT_{\m\n}^{_{(4)}} \neq 0$. Therefore, establishing if the route A or the route B (or none of the two) gives the correct result in the pure tension case gives an invaluable indication about the correct way to perform the pillbox integration in the general case.

As we show in appendix \ref{Pure tension perturbations}, at linear level in $\dla$ the  normal component of the bending $\hdvfn^{[n]}$ in the pure tension perturbation case reads
\begin{align}
\label{heyhey}
\hdvfn^{[n]}(\hxi) &\simeq \cos \bigg( \frac{\bla}{2 M_6^4} \, \ep_{[n]} (\hxi) \bigg) \, \d\!\b^{(4)}_{[n]} + \frac{\dla}{2 M_6^4} \, \cos \bigg( \frac{\bla}{2 M_6^4} \, \ep_{[n]} (\hxi) \bigg) \, \int_{0}^{\hxi} \! d \z \, \, \ep_{[n]} (\z) \, \cos \bigg( \frac{\bla}{2 M_6^4} \, \ep_{[n]} (\z) \bigg) + \nn \\[2mm]
& + \frac{\dla}{2 M_6^4} \, \sin \bigg( \frac{\bla}{2 M_6^4} \, \ep_{[n]} (\hxi) \bigg) \, \int_{0}^{\hxi} \! d \z \, \, \ep_{[n]} (\z) \, \sin \bigg( \frac{\bla}{2 M_6^4} \, \ep_{[n]} (\z) \bigg)
\end{align}
and in particular its value on the side of the cod-2 brane reads
\begin{align}
\label{enoughisenough}
\hdvfn^{[n]}\Big\rvert_{l_{2}^{[n]}} &\simeq \cos \bigg( \frac{\bla}{4 M_6^4} \bigg) \, \d\!\b^{(4)}_{[n]} + \frac{\dla}{2 M_6^4} \, \cos \bigg( \frac{\bla}{4 M_6^4} \bigg) \, \int_{0}^{l_{2}^{[n]}} \! d \z \, \, \ep_{[n]} (\z) \, \cos \bigg( \frac{\bla}{2 M_6^4} \, \ep_{[n]} (\z) \bigg) + \nn \\[2mm]
& + \frac{\dla}{2 M_6^4} \, \sin \bigg( \frac{\bla}{4 M_6^4} \bigg) \, \int_{0}^{l_{2}^{[n]}} \! d \z \, \, \ep_{[n]} (\z) \, \sin \bigg( \frac{\bla}{2 M_6^4} \, \ep_{[n]} (\z) \bigg)
\end{align}
where $\dla$ is the tension perturbation and the \emph{regulating function} $\ep_{[n]}$ is defined as
\beq
\ep_{[n]}(\hxi) = \int_{0}^{\hxi} f_{1}^{[n]}(\z) \, d \z
\eeq
Note that $\d\!\b^{(4)}_{[n]}$ in this case is independent of the 4D coordinates, as a consequence of the translational invariance in the 4D directions which is enjoyed by the pure tension solutions, and so is truly a number. Our aim is then to compute numerically the integral
\beq
\label{Alkistis}
\mscr{I}_{[n]} = \frac{\bla}{2 M_6^4} \int_{-l_{2}^{[n]}}^{+l_{2}^{[n]}} \! d \hxi \,\, f_{1}^{[n]}(\hxi) \,\, \hdvfn^{[n]}(\hxi)
\eeq
and to compare the result with the value
\beq
\label{routeAdef}
\mscr{A}_{[n]} = 2 \, \tan \bigg( \frac{\bla}{4 M_6^4} \bigg) \, \hdvfn^{[n]}\Big\rvert_{l_{2}^{[n]}}
\eeq
and with the value
\beq
\label{routeBdef}
\mscr{B}_{[n]} = \frac{\bla}{2 M_6^4} \,\, \hdvfn^{[n]}\Big\rvert_{l_{2}^{[n]}}
\eeq
The routes A and B in fact claim that the result of the pillbox integration $\mscr{I}$ of (\ref{Vittoria}) and of (\ref{Vittoria2}) is respectively the limit for $n \to \infty$ of the sequences $\mscr{A}_{[n]}$ and $\mscr{B}_{[n]}$. Since $\mscr{I}$ is the limit of $\mscr{I}_{[n]}$ for $n \to \infty$, we conclude that, if the route A is correct, the sequence $\mscr{A}_{[n]}$ and the sequence $\mscr{I}_{[n]}$ have to converge for $n \to \infty$ 
\beq
\label{routeA}
\lim_{n \rightarrow + \infty} \mscr{I}_{[n]} = \lim_{n \rightarrow + \infty} \mscr{A}_{[n]}
\eeq
while, if the route B is correct, the sequence $\mscr{B}_{[n]}$ and the sequence $\mscr{I}_{[n]}$ have to converge
\beq
\label{routeB}
\lim_{n \rightarrow + \infty} \mscr{I}_{[n]} = \lim_{n \rightarrow + \infty} \mscr{B}_{[n]}
\eeq

\subsubsection{Choice of the free parameters}
\label{Choice of the free parameters}

The normal component of the bending is expressed in terms of the quantities $\dla$, $\ep_{^{[n]}}(\hxi)$ and $\d\!\b^{(4)}_{^{[n]}}$, which play the role of free parameters. The first one fixes the amplitude of the tension perturbation, and is indeed a free parameter apart from the fact that it has to satisfy the condition $\dla / \bla \ll 1$. The regulating function $\ep_{^{[n]}}(\hxi)$, instead, expresses the details of the internal structure of the cod-2 brane and is therefore fixed once we choose the system we are working with. For the purpose of checking numerically the validity of route A and B, it is enough to choose a particular realization of $\ep_{^{[n]}}$ and $f_{1}^{_{[n]}}$: we use the following realization of the Dirac delta
\beq
\label{fnexplicit}
f_{1}^{[n]}(\hxi) = 
\begin{cases}
\dfrac{n}{2 \pi} \, \Big( 1 + \cos\big( n \, \hxi \big) \Big) & \text{for $\abs{\hxi} \leq \dfrac{\pi}{n}$}
\vspace{3mm} \\
0 & \text{for $\abs{\hxi} > \dfrac{\pi}{n}$}
\end{cases}
\eeq
and the associated regulating function
\beq
\label{epnexplicit}
\ep_{[n]}(\hxi) =
\begin{cases}
\dfrac{1}{2 \pi} \, \Big( n \, \hxi + \sin \big( n \, \hxi \big) \Big) & \text{for $\abs{\hxi} \leq \dfrac{\pi}{n}$} \vspace{3mm} \\
\pm \dfrac{1}{2} & \text{for $\hxi \gtrless \pm \, \dfrac{\pi}{n}$}
\end{cases}
\eeq
\begin{figure}[htp!]
\begin{center}
\includegraphics{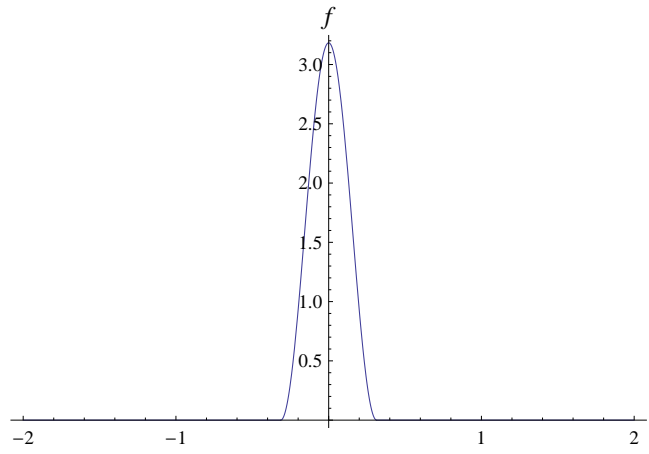}
\caption[The realization $f_1$ of the Dirac delta]{The realization $f_1$ of the Dirac delta}
\label{fdelta}
\end{center}
\end{figure}
\begin{figure}[htp!]
\begin{center}
\includegraphics{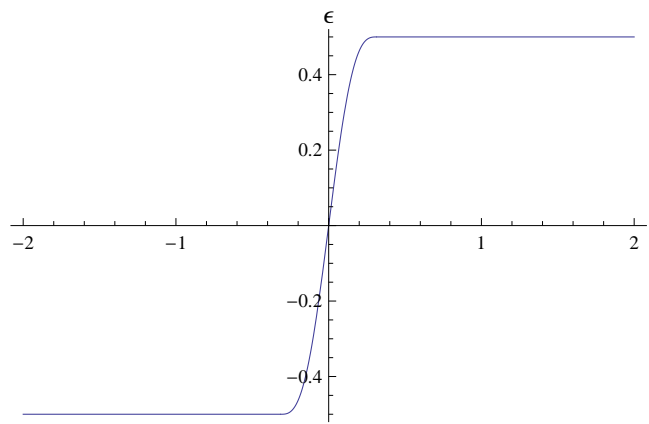}
\caption[The regulating function $\ep$]{The regulating function $\ep$}
\label{epsilon}
\end{center}
\end{figure}
whose plots for $n = 10$ are shown respectively in figure \ref{fdelta} and in figure \ref{epsilon}. Note that in this case the thickness of the (physical) cod-2 brane is $l_2 = \pi/n$, and indeed the thin limit $l_2 \rightarrow 0^+$ mathematically corresponds to the limit $n \rightarrow + \infty$. It is worthwhile to point out that the explicit form (\ref{fnexplicit}) for the function $f_{1}^{_{[n]}}$ is of class $\mscr{C}^1$ on all the real axis, but its second derivative does not exist in $\hxi = \pm \pi/n$. However, this is not a problem for what concerns the numerical check since the latter does not involve the derivation of the function $f_{[n]}$ but only its integration.

The quantity $\d\!\b^{_{(4)}}_{^{[n]}}$, instead, is in general determined by the equations of motion once we specify the source configuration. However, in the pure tension case $\d\!\b^{_{(4)}}_{^{[n]}}$ is not a 4D field but a number (as we mentioned above), and its value is not fixed by the equations of motion since its 4D D'Alembertian $\boxf \d\!\b^{_{(4)}}_{^{[n]}}$ (which is the quantity which appears in the junction conditions) vanishes identically. This is consistent with the fact that a rigid translation of the cod-1 and cod-2 branes is a symmetry of the system, since the bulk metric is the 6D Minkowski metric. However, our aim here is to understand which route (A or B, or none of the two) to evaluate the integral $\mscr{I}$ is correct from a mathematical point of view, independently of the fact that the integral itself does or does not contribute to the equations of motion. Therefore, in the particular case we are considering, $\d\!\b^{_{(4)}}_{^{[n]}}$ can be considered a free parameter as well.

To test the validity of the routes A and B, it is convenient to choose the free parameters in such a way that the numbers $\mscr{A}_{[n]}$ and $\mscr{B}_{[n]}$ are quite different. Taking a look at (\ref{routeAdef}) and (\ref{routeBdef}) it is clear that we should avoid the choice $\bla/4 M_6^4 \ll 1$, since the difference between $\mscr{A}_{[n]}$ and $\mscr{B}_{[n]}$ increases as $\bla/4 M_6^4$ increases. It is therefore useful to choose the background tension close to the maximum tension. Moreover, we should choose $\d\!\b^{_{(4)}}_{^{[n]}}$ in such a way that $\hdvfn^{_{[n]}}$ is not too small at the side of the ribbon brane. These considerations prompt us to choose the background tension and the tension perturbation to be
\begin{align}
\label{FraMarche}
\bla &= \frac{3}{4} \, \bla_M & \frac{\dla}{2 \Msf} &= 0.1
\end{align}
which is consistent with the hypothesis that the tension perturbation is small since with this choice we have $\dla/\bla \simeq 0.04$ . Furthermore, for $\d\!\b^{_{(4)}}_{^{[n]}}$ we choose the value $\d\!\b^{_{(4)}}_{^{[n]}} = 5$.

\subsubsection{Numerical results for the pillbox integration}

Having chosen a specific realization of the internal structure $f_{1}^{_{[n]}}$ and $\ep_{^{[n]}}$, and having fixed the free parameters, we can evaluate numerically the integral $\mscr{I}_{[n]}$ for several values of $n$.

\begin{figure}[htp!]
\begin{center}
\includegraphics{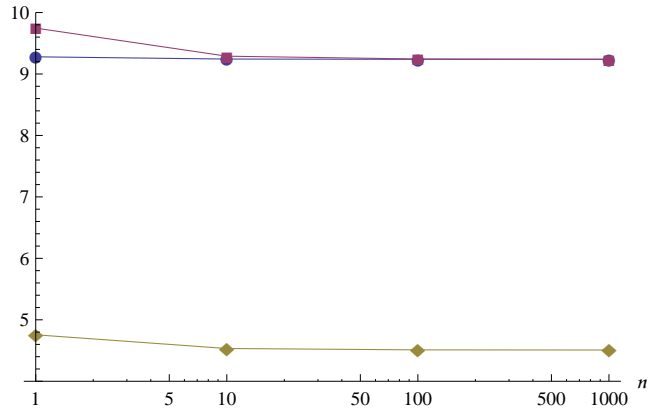}
\caption[Numerical results of the pillbox integration]{Plot of the numerical results of the pillbox integration}
\label{PillboxIntegrationfigure}
\end{center}
\end{figure}

The results of the numerical integrations are given in table \ref{pillboxintegrationtable} with $5$ significant digits, and for clarity the same results are plotted in figure \ref{PillboxIntegrationfigure} (note that the plot is semi-logaritmic). It is evident that the points corresponding to $\mscr{A}_{[n]}$ (squares) converge to the points corresponding to $\mscr{I}_{[n]}$ (circles), while the points corresponding to $\mscr{B}_{[n]}$ (diamonds) are significantly distant from the former ones. 
\begin{table}[htb!]
\centering
\[
\begin{array}{ccccc}
\toprule
n & 1 & 10 & 10^2 & 10^3 \\
\midrule
\mscr{I}_{[n]} & 9.2794 & 9.2429 & 9.2392 & 9.2388 \\
\mscr{A}_{[n]} & 9.7466 & 9.2896 & 9.2439 & 9.2393 \\
\mscr{B}_{[n]} & 4.7562 & 4.5332 & 4.5109 & 4.5086 \\
\bottomrule
\end{array}
\]
\caption{Numerical results of the pillbox integration}
\label{pillboxintegrationtable}
\end{table}
This implies that, at least in the pure tension perturbation case, the route B is wrong while the route A is correct. In particular, the pillbox integration performed following the route B gives a lower value compared to the pillbox integration performed following the route A because the route B completely misses the peak of $\hdvfn^{[n]}$ inside the cod-2 brane (see figure \ref{normalbending}).

The same conclusion can be reached in a slightly different way, by exploiting the fact that $\d\!\b^{_{(4)}}_{^{[n]}}$ and $\dla$ are independent parameters. In fact, both $\mscr{I}_{^{[n]}}$ and $\hdvfn^{_{[n]}}$ are the sum of a piece multiplied by $\d\!\b^{_{(4)}}_{^{[n]}}$ (which we call the ``bending piece'') and a piece multiplied by $\dla$ (which we call the ``tension piece''); since these two parameters are independent, if one of the equations (\ref{routeA}) and (\ref{routeB}) is valid then it has to be valid also separately for the bending piece and for the tension piece. Note that the bending piece of $\hdvfn^{_{[n]}}\big\rvert_{l_{2}^{_{[n]}}}$ reads
\beq
\label{bendingN}
\text{bending} \bigg[ \hdvfn^{[n]}\Big\rvert_{l_{2}^{[n]}} \bigg] = \cos \bigg( \frac{\bla}{4 M_6^4} \bigg)
\eeq
while the bending piece of $\mscr{I}_{[n]}$ reads
\beq
\text{bending} \Big[ \mscr{I}_{[n]} \Big] = \frac{\bla}{2 M_6^4} \int_{-l_{2}^{[n]}}^{+l_{2}^{[n]}} \! d \hxi \,\, f_{[n]}(\hxi) \,\, \cos \bigg( \frac{\bla}{2 M_6^4} \, \ep_{[n]} (\hxi) \bigg)
\eeq
The latter integral can be performed exactly changing the integration variable to $\z = \ep_{[n]} (\hxi)$, to obtain
\beq
\label{integralexact}
\text{bending} \Big[ \mscr{I}_{[n]} \Big] = 2 \, \sin \bigg( \frac{\bla}{4 M_6^4} \bigg)
\eeq
Putting together the formulas (\ref{bendingN}) and (\ref{integralexact}) we reproduce exactly the result of route A
\beq
\text{bending} \Big[ \mscr{I}_{[n]} \Big] = 2 \, \tan \bigg( \frac{\bla}{4 M_6^4} \bigg) \,\, \text{bending} \bigg[ \hdvfn^{[n]}\Big\rvert_{l_{2}^{[n]}} \bigg]
\eeq

\subsubsection{Discussion}

The analysis of the sections \ref{Ghost-free regions in parameters space} and \ref{Numerical check} confirms the suggestion that the difference between the two results (\ref{critical tension}) and (\ref{critical tension theirs}) comes from the pillbox integration. It is in fact very unexpected that an a priori innocent procedure, such as exploiting the presence of a Dirac delta in the equations, is not justified in the nested-branes with induced gravity set-ups. This is a consequence of the fact that the singular structure of the geometry is very subtle, and indirectly confirms the belief that the singular structure of branes of codimension higher than one is in general more complex than the singular structure of codimension-1 branes. In fact, while in the codimension-1 case there is essentially a unique way to render the brane thin, in higher codimension set-ups there is an infinite number of non-equivalent ways to do that. For example, in our regularization choice the cod-2 brane is ``stripe-like'', so in some sense it is a codimension-1 regularization of a codimension-2 object, but we could have considered as well a circular or a cross-like regularization (which is probably best suited when the cod-2 brane lies at the intersection of two cod-1 branes), or many others. Each of these regularization choices has its own peculiar singular structure, and the equations which describe the thin limit behaviour of the system have to be derived independently for each case taking great care of its peculiarities.

This analysis puts on firm footing our derivation of the thin limit equations for the nested-branes realization of the 6D Cascading DGP. Concerning the critical tension, it strongly supports our claim that the correct value of the critical tension is (\ref{critical tension}), and that also models where gravity displays a direct transition 6D $\to$ 4D ($m_6 \gg m_5$) are phenomenologically viable. It is interesting to speculate that also the $m_6 \gg m_5$ region of the parameter space may still lead to a cascading behaviour. In fact, in this case the critical tension is very close to the maximal tension (so the deficit angle is close to $2 \pi$). In this regime, the angular direction around the codimension 2 source is quasi-compactified, resulting in an intermediate 5D behaviour \cite{Kaloper:2007ap, Kaloper:2007qh}. We don't pursue this point further in this paper, but it leaves open the possibility that our result is compatible with the spirit of \cite{deRham:2007xp}.

It is worthwhile to note that the use of the bulk-based approach has been fundamental in our analysis to identify clearly the difference in the singular behaviour between different perturbation fields (bending and metric). In fact, in a brane-based approach every information is encoded in the bulk metric perturbations, and (although possible) we feel that it is less intuitive to distinguish the converge properties of the different components of the metric. Although we already showed above that to reproduce the result of \cite{deRham:2010rw} in our approach we need to use a procedure which is not mathematically justified, it is interesting to see where the subtlety lies directly in the approach of \cite{deRham:2010rw}. In their analysis, the 4D Einstein-Hilbert action receives a contribution from the integration across the cod-2 brane of the 5D Einstein-Hilbert action and of the 5D boundary effective action induced on the cod-1 brane by the 6D Einstein-Hilbert action. The 4D Einstein-Hilbert action generates the $\bla$-independent ghostly 4D kinetic term for the field $\pi$, while the $\bla$-dependent healthy kinetic term on the cod-2 brane is generated by the integration across the cod-2 brane of the mixed term
\beq
\frac{3}{2} \, \frac{\bN_y^{\p}}{\bN^2} \, \pi \, \boxf \, \ti{\s}
\eeq
where $\bN_y$ and $\bN$ are defined in appendix \ref{The brane-based approach} and $\ti{\s}$ is a perturbation field related to the perturbation of the shift vector of the cod-1 brane. Here a prime stands for derivation with respect to $y$, which is the coordinate on the cod-1 brane which is normal to the cod-2 brane, and the background relation (\ref{Kelly}) ensures that in the thin limit
\beq
\frac{\bN_y^{\p}}{\bN} \sim \frac{\bla}{2 M_6^4} \, \d(y)
\eeq
The pillbox integration in \cite{deRham:2010rw} is executed as follows
\beq
\label{pillboxinttheirs}
\int \frac{3}{2} \, \frac{\bN_y^{\p}}{\bN^2} \, \pi \, \boxf \, \ti{\s} \, dy \sim \int \frac{3}{2} \, \frac{1}{\bN} \, \pi \, \boxf \, \ti{\s} \, \frac{\bla}{2 M_6^4} \, \d(y) \, dy \sim \frac{3 \bla}{4 M_6^4} \, \frac{1}{\bN} \, \pi \, \boxf \, \ti{\s} \Big\rvert_{y = 0}
\eeq
and upon de-mixing the fields $\pi$ and $\ti{\s}$ this term gives the healthy kinetic term for $\pi$ on the cod-2 brane. However, a careful analysis shows that the field $\bN$ has a non-trivial behaviour near the cod-2 brane, namely it is peaked at $y = 0$ as much as $\dvfn$ is in our analysis, and that also the field $\ti{\s}$ has a non-trivial behaviour there. For the same reason explained in section \ref{Pillbox integration and convergence properties}, the properties of the Dirac delta cannot be used in evaluating the integral (\ref{pillboxinttheirs}), and so the derivation is not mathematically justified.

\section{Conclusions}
\label{Conclusions}

In this paper we studied the behaviour of weak gravitational fields in the 6D Cascading DGP model, with the aim of understanding geometrically why a critical tension emerges in the model. We considered a specific realization of the Cascading DGP, which we called the nested brane realization, where the codimension-1 brane can be considered thin with respect to the codimension-2 brane. We considered solutions which correspond to pure tension sources on the codimension-2 brane, and studied perturbations of the bulk geometry and of the embedding of the codimension-1 brane at first order around these background solutions. We performed a 4D scalar-vector-tensor decomposition of the perturbation fields, and focused on the scalar sector, which has been shown to be the only relevant sector concerning the critical tension.

We showed that the master variable of the scalar sector obeys a master equation on the cod-2 brane when the latter is thin, and that in a suitable 4D limit its dynamics on the codimension-2 brane decouples from its dynamics on the codimension-1 brane and in the bulk. The decoupled equation suggests that the master variable is an effective 4D ghost when the background tension is smaller than a critical value $\blac$, while it is a healthy field otherwise. We gave a geometrical interpretation of why the value of the background tension influences the fact that the master variable is a ghost or not, and how the critical tension emerges. The value of the critical tension in our analysis is however different from the value found in the literature. This difference has an important implication because, contrary to the claim in the literature, our result implies that, for every value of the free parameters of the theory, there exist values of the background tension such that the model is ghost free. In particular, also the models where the behaviour of gravity undergoes a direct 6D $\to$ 4D transition when we move from large to small scales are phenomenologically viable.

We identified the source of this difference with the way the singularity at the codimension-2 brane is taken care of, and in particular with the procedure used to perform the pillbox integration across the codimension-2 brane. We showed that the result in the literature relies on the use of the properties of the Dirac delta which is however not justified in these set-ups, due to the subtlety of the singular structure of the fields inside the cod-2 brane. To provide an independent test of the validity of the two results, we performed the pillbox integration numerically in a particular case where the solution for the perturbation fields inside the codimension-2 brane is known exactly. The outcome agrees with our result, which supports our analysis and the value of the critical tension that we found. We stress that the existence of the induced gravity term on the codimension-1 brane is crucial to avoid ghosts. This is manifest in our geometrical interpretation, and can be seen also by considering the limit $\Mft \to 0$ at $\Msf$ and $\Mfs$ fixed. In this limit, the critical tension tends to the maximum tension $\blac \to \bla_M$ and therefore the presence of the ghost is inevitable. This is compatible with the findings of \cite{Dubovsky:2002jm} that the 6D (cod-2) DGP model has a perturbative ghost around flat solutions.

\section*{Acknowledgments}
FS wishes to thank Paolo Creminelli for hospitality at the Abdus Salam's ICTP, Trieste, Italy where part of this work has been done. FS and KK were supported by the European Research Council's starting grant. KK is supported by the UK Science and Technology Facilities Council grants number ST/K00090/1 and ST/L005573/1.

\newpage

\begin{appendix}

\section{The brane-based approach}
\label{The brane-based approach}

In this appendix we describe the analysis in the brane-based approach performed in \cite{deRham:2010rw}, and its relationship with our analysis. Note that the conventions we use in the rest of the papers does not hold in this appendix.

\subsection{Pure tension solutions from a brane-based point of view}

In \cite{deRham:2010rw}, pure tension solutions in the Cascading DGP model were derived. In those solutions, a 6D spacetime is covered by a coordinate chart $(\bar{z}, \bar{y},\bar{x}^{\cdot})$, where $\bar{z}$ is defined on $(0, +\infty)$ and $(\bar{y},\bar{x}^{\cdot})$ are defined on $\mathbb{R}^5$, and the bulk geometry is defined by the line element
\beq
\label{backgroundthemlineapp}
ds^2 = \big( 1 + \b^2 \big) \, d\bar{z}^2 + 2 \b \, \bep(\bar{y}) \, d\bar{z} d\bar{y} + d\bar{y}^2 + \e_{\m \n} d\bar{x}^{\m} d\bar{x}^{\n}
\eeq
where $\b$ is a real parameter and $\bep(\bar{y})$ is a smooth function which is a regularized version of the ``symmetric'' step function. In particular, $\bep$ is odd with respect to the reflection $\bar{y} \to -\bar{y}$ and asymptotes the value $\pm 1$ when $\bar{y} \to \pm \infty$, so the convention is different from the one used in the main text for $\ep(y)$ (which asymptotes $\pm 1/2$). Let's suppose that a (thin) cod-1 brane is placed at $\bar{z} = 0$, and let's choose to parametrize it with the bulk coordinates $(\bar{y},\bar{x}^{\cdot})$, and that a (mathematical) cod-2 brane is placed at $\bar{z} = \bar{y} = 0$, and let's choose to parametrize it with the bulk coordinates $(\bar{x}^{\cdot})$. It is not difficult to see that the 6D Riemann tensor built from the metric (\ref{backgroundthemlineapp}) vanishes identically, and that the induced metrics on the cod-1 and cod-2 branes are respectively the 5D and the 4D Minkowski metrics (independently of the explicit form of $\bep$), so in particular the metric (\ref{backgroundthemlineapp}) satisfies the bulk Einstein equations. However, the extrinsic geometry of the cod-1 brane is non-trivial, since we have\footnote{Here a prime $\phantom{i}^{\p}$ indicates a derivative with respect to $\bar{y}$.}
\begin{align}
\label{extrinsic curvature dRKT}
\bar{K}_{\m \n} &= 0 & \bar{K}_{\m y} &= 0 & \bar{K}_{yy} &= - \, \beta \, \dfrac{\bep^{\, \prime}(\bar{y})}{\sqrt{ 1 + \b^2 \big( 1 - \bep (\bar{y})^2 \big)}} 
\end{align}
Suppose now that the cod-1 brane contains a pure tension source localized around the cod-2 brane, so that the energy-momentum on the cod-1 brane is of the form
\beq
\bar{T}_{ab} = - \bla \, \bar{f}(\bar{y}) \,\, \d_{a}^{\,\,\,\m} \, \d_{b}^{\,\,\,\n} \,\, \e_{\m\n}
\eeq
where $\bar{f}(\bar{y})$ is a positive, even and normalized function (so it is a regularized version of the Dirac delta function) which describes the details of the distribution of the tension inside a thick cod-2 brane whose boundaries are $\bar{y} = \pm l_2$. The only non-trivial component of the junction conditions reads
\beq
\label{Kelly}
\frac{\beta \, \bep^{\, \prime}(\bar{y})}{\sqrt{ 1 + \b^2 \big( 1- \bep^2(\bar{y}) \big)}} = \frac{\bla}{2 M_6^4} \, \bar{f}(\bar{y})
\eeq
which in particular implies that $\bep(\bar{y}) = \pm 1$ for $\bar{y} \gtrless \pm l_2$. If $\bla < 2 \pi \Msf$ the equation above admits global solutions, and integrating it over the interval $[-l_2 , l_2]$ we obtain
\beq
\label{Julie}
\arctan \b = \frac{\bla}{4 M_6^4}
\eeq
which implies that for $\abs{\bar{y}} > l_2$ the metric (\ref{backgroundthemlineapp}) is fixed by the total amount of tension $\bla$ present inside the thick cod-2 brane, while for $\abs{\bar{y}} < l_2$ the shape of $\bep(\bar{y})$ explicitly depends on the details of how the tension is distributed inside the thick cod-2 brane. In the thin limit, defined by $l_2 \rightarrow 0^+$ keeping $\bla$ constant, $\b$ remains constant while $\bar{f}$ tends to a Dirac delta and $\bep$ tends to the (symmetric) step function.

Note that in this case the embedding of cod-1 brane is straight, even in the thin limit. However, the bulk metric becomes discontinuous in the thin limit ($\bar{g}_{zy} = \b \, \bep(\bar{y})$): this is necessary to generate a delta function divergence in the extrinsic curvature (cfr.~(\ref{extrinsic curvature dRKT}) and (\ref{Kelly})), as we mentioned in section \ref{Small perturbations in the bulk-based approach}.

\subsection{Equivalence with the bulk-based description}

The geometry of the bulk-branes system corresponding to the metric (\ref{backgroundthemlineapp}) is however not evident. The fact that the Riemann tensor is identically vanishing in the bulk implies that (\ref{backgroundthemlineapp}) is equivalent to a portion of a 6D Minkowski space written in a non-trivial coordinate system. To have a transparent idea of the geometry of the configuration (\ref{backgroundthemlineapp}), we can try to find a coordinate transformation which maps it into the 6D Minkowski space. The geometrical meaning of the configuration will then be encoded in the embedding of the cod-1 brane, which after the coordinate change will be non-trivial.

\subsubsection{The change of coordinates}
\label{The change of coordinates}

Let's start from the configuration (\ref{backgroundthemlineapp})
\begin{equation*}
\overline{g}_{zz} = 1 + \b^2 \quad \qquad \overline{g}_{zy} = \b \, \bep (\bar{y}) \quad \qquad \overline{g}_{yy} = 1 \quad \qquad \overline{g}_{z \m} = \overline{g}_{y \m} = 0 \quad \qquad \overline{g}_{\m\n} = \e_{\m\n} 
\end{equation*}
and consider the following coordinate transformation
\begin{equation*}
(\star) \quad \left\{
\begin{aligned}
  \bar{z}(\hat{z}, \hat{y}, \hat{x}^{\cdot}) &= \frac{1}{\sqrt{1 + \bq}} \, \Big( \hat{z} - \mathscr{F}(\hat{y}) \Big) \\
  \bar{y}(\hat{z}, \hat{y}, \hat{x}^{\cdot}) &= \hat{y}\\[3mm]
  \bar{x}^{\mu}(\hat{z}, \hat{y}, \hat{x}^{\cdot}) &= \hat{x}^{\mu}
\end{aligned}
\right.
\end{equation*}
which brings the metric into the form
\begin{gather*}
\hat{g}_{zz} = 1 \qquad \qquad \hat{g}_{zy} = - \frac{d\mathscr{F}}{d \hat{y}} \, + \, \frac{\b \, \bep (\hat{y})}{\sqrt{1+ \b^2}} \qquad \qquad \hat{g}_{yy} = \Big( \frac{d\mathscr{F}}{d \hat{y}} \Big)^{\!2} - 2 \, \frac{d\mathscr{F}}{d \hat{y}} \, \frac{\b \, \bep (\hat{y})}{\sqrt{1+ \b^2}} + 1 \notag \\
\\
\hat{g}_{z \m} = \hat{g}_{y \m} = 0 \qquad \qquad \qquad \qquad \qquad \qquad \hat{g}_{\m\n} = \e_{\m\n} \notag
\end{gather*}
Asking that $\hat{g}_{zy} = 0$ amounts to impose
\beq
\label{Fuji}
\frac{d\mathscr{F}}{d \hat{y}} (\hat{y}) = \frac{\b \, \bep (\hat{y})}{\sqrt{1+ \b^2}}
\eeq
which in turn implies
\begin{gather*}
\hat{g}_{zz} = 1 \quad \qquad \hat{g}_{zy} = 0 \quad \qquad \hat{g}_{yy} = 1 - \, \Bigg( \!\frac{\b \, \bep (\hat{y})}{\sqrt{1+ \b^2}} \! \Bigg)^{\!\!2} \quad \qquad \hat{g}_{z \m} = \hat{g}_{y \m} = 0 \quad \qquad \hat{g}_{\m\n} = \e_{\m\n}
\end{gather*}
Secondly, consider the following coordinate transformation
\begin{equation*}
(\star \star) \quad \left\{
\begin{aligned}
  \hat{z}(z, y, x^{\cdot}) &= z \\
  \hat{y}(z, y, x^{\cdot}) &= \mathscr{G}(y)\\
  \hat{x}^{\mu}(z, y, x^{\cdot}) &= x^{\mu}
\end{aligned}
\right.
\end{equation*}
which brings the metric into the form:
\begin{gather*}
g_{zz} = 1 \qquad \qquad g_{zy} = 0 \qquad \qquad g_{yy} = \, \Bigg( \frac{d\mathscr{G}}{d y} \Bigg)^{\!\!2} \, \Bigg[ 1 - \Bigg( \!\frac{\b \, \bep (\mathscr{G}(y))}{\sqrt{1+ \b^2}} \! \Bigg)^{\!\!2} \, \Bigg] \notag \\
\\
g_{z \m} = g_{y \m} = 0 \qquad \qquad \qquad \qquad \qquad \qquad g_{\m\n} = \e_{\m\n} \notag
\end{gather*}
Asking that $g_{yy} = 1$ amounts to
\beq
\label{Lex}
\Bigg( \frac{d\mathscr{G}}{d y} \Bigg)^{\!\!2} =  \frac{1+ \b^2}{1+ \b^{2} \, \Big( 1 - \bep^{2} \big( \mathscr{G}(y) \big) \Big)}
\eeq
which implies
\beq
\label{voila}
g_{AB} = \e_{AB}
\eeq
Therefore, provided that the functions $\mathscr{F}$ and $\mathscr{G}$ exist, the composition of the two coordinates changes ($\star$) and ($\star\star$) transforms the initial metric (\ref{backgroundthemlineapp}) into the 6D Minkowski metric. The existence of solutions of the differential equation (\ref{Fuji}) is ensured by the fact that the function $\bep$, being continuous, is primitivable. Concerning the existence of the function $\mscr{G}$, note first of all that the right hand side of (\ref{Lex}) never vanishes, so there are two classes of solutions characterised by the fact that $d\mathscr{G}/dy$ is positive or negative. These two choices for the sign of $d\mathscr{G}/dy$ correspond to the fact that the new ``$y$'' coordinate ($y$) has the same or the opposite orientation with respect to the old ``$y$'' coordinate ($\bar{y}$): we choose to impose that $d\mathscr{G}/dy$ is positive, which means that the $y$ coordinate has the same orientation as $\bar{y}$. Therefore, we can rewrite the equation (\ref{Lex}) as
\beq
\label{Lex2}
\frac{d\mathscr{G}}{d y} =  \mathcal{D} \big( \mathscr{G}(y) \big)
\eeq
where
\beq
\label{Lex3}
\mathcal{D} \big( \mathscr{G} \big) = \sqrt{\frac{1+ \b^2}{1+ \b^{2} \, \Big( 1 - \bep^{2} \big( \mathscr{G} \big) \Big)}}
\eeq
Since both $\bep$ and $\bep^{\, \p}$ are smooth and bounded by hypothesis, the function $\mcal{D}$ is (globally) Lipschitzian: therefore, the Picard-Lindel\"of theorem (see for example \cite{Hale}) ensures that, for each choice of the initial condition, there exists a unique local solution to the equation (\ref{Lex2}). Furthermore, the fact that $\mcal{D}$ is smooth and bounded both from below and from above (we have in fact $1 \leq \mathcal{D} ( \mathscr{G} ) \leq \sqrt{1 + \bq}\,$) implies that the local solution can be extended to a global solution. Moreover, it also implies that $\mathscr{G}(y)$ is a diffeomorphism $\mathbb{R} \to \mathbb{R}$ and in particular is invertible.

Therefore, we can indeed find a change of coordinates which maps the metric (\ref{backgroundthemlineapp}) into the 6D Minkowski metric: in the new reference system, the geometrical meaning of the configuration is encoded in the trajectory of the cod-1 brane, which is defined by $\mscr{F}$ and $\mscr{G}$. In synthesis, we have passed from a trivial embedding and a non-trivial metric to a non-trivial embedding and a trivial metric.

\subsubsection{The new embedding of the cod-1 brane}

To find the embedding of the cod-1 brane in the new bulk reference system, note first of all that we can still parametrize the cod-1 brane and the cod-2 brane with the ``old'' coordinates $(\bar{y}, \bar{x}^{\cdot})$ and $\bar{x}^{\cdot}$. Furthermore, as a consequence of the two coordinate changes, a point $(\bar{z},\bar{y},\bar{x}^{\cdot}) = (0,\bar{y},\bar{x}^{\cdot})$ on the cod-1 brane is mapped into the point $(z,y,\xd) = (\mathscr{F}(\bar{y}),\mathscr{G}^{-1}(\bar{y}),\bar{x}^{\cdot})$, and in particular a point $(\bar{z},\bar{y},\bar{x}^{\cdot}) = (0,0,\bar{x}^{\cdot})$ on the cod-2 brane is mapped into the point $(z,y,\xd) = (\mathscr{F}(0),\mathscr{G}^{-1}(0),\bar{x}^{\cdot})$. Therefore, the embedding of the cod-1 brane into the 6D Minkowski space is then
\beq
\label{Spain}
\vf^A(\bar{y},\bar{x}^{\cdot}) = \big( \mcal{Z}(\bar{y}),\mcal{Y}(\bar{y}), \bar{x}^{\cdot} \big)
\eeq
where $\mcal{Z}(\bar{y}) \equiv \mscr{F}(\bar{y})$ and $\mcal{Y}(\bar{y}) \equiv \mscr{G}^{-1}(\bar{y})$. Note that, as a consequence of (\ref{Fuji}) and (\ref{Lex}), the components of the embedding function $\mcal{Z}$ and $\mcal{Y}$ satisfy
\beq
\label{relation a}
\mcalZpq (\bar{y}) + \mcalYpq (\bar{y}) = 1
\eeq

The components of the embedding function (\ref{Spain}) are not uniquely determined by the differential equations (\ref{Fuji}) and (\ref{Lex}), since to determine them we need to add some initial conditions. We choose to impose that the position of the cod-2 brane have the same bulk coordinates before and after the coordinate changes, which means to ask that $\mscr{F}(0) = 0$ and $\mscr{G}(0) = 0$. The non-trivial components of the embedding function of the cod-1 brane are then determined by the following Cauchy problems
\beq
\label{mcalZ Cauchy problem}
\left\{
\begin{aligned}
\mcal{Z}^{\p}(\bar{y}) &= \frac{\b \, \bep (\bar{y})}{\sqrt{1+ \b^2}} \\[2mm]
\mcal{Z}(0) &= 0 
\end{aligned}
\right.
\eeq
and
\beq
\label{mcalY Cauchy problem}
\left\{
\begin{aligned}
\mcal{Y}^{\p}(\bar{y}) &= \sqrt{\frac{1 + \bq \big( 1 - \bep^2(\bar{y}) \big)}{1 + \bq}} \\[2mm]
\mcal{Y}(0) &= 0
\end{aligned}
\right.
\eeq 

\noi On the other hand, the relation (\ref{relation a}) implies that there exists a function $\mcal{S}(\bar{y})$ such that
\begin{align}
\label{grumpy}
\mcal{Z}^{\p}(\bar{y}) &= \sin \mcal{S}(\bar{y}) & \mcal{Y}^{\p}(\bar{y}) &= \cos \mcal{S} (\bar{y})
\end{align}
which has to be odd to respect the parity of $\mcal{Z}^{\p}$ and $\mcal{Y}^{\p}$. The Cauchy problems (\ref{mcalZ Cauchy problem}) and (\ref{mcalY Cauchy problem}) then imply that the derivative of $\mcal{S}$ reads
\beq
\mcal{S}^{\p}(\bar{y}) = \frac{d}{d \bar{y}} \, \arctan \bigg( \frac{\mcal{Z}^{\p}(\bar{y})}{\mcal{Y}^{\p}(\bar{y})} \bigg) = \frac{\b \, \bep^{\, \p} (\bar{y})}{\sqrt{1 + \bq \big( 1 - \bep^2(\bar{y}) \big)}}
\eeq
and so, taking into account (\ref{Kelly}), $\mcal{S}$ is determined by the Cauchy problem
\beq
\label{mcalS Cauchy problem}
\left\{
\begin{aligned}
\mcal{S}^{\p}(\bar{y}) &= \frac{\bla}{2 \Msf} \, \bar{f}(\bar{y}) \\[2mm]
\mcal{S}(0) &= 0
\end{aligned}
\right.
\eeq
Taking a look at equations (\ref{voila}), (\ref{Spain}), (\ref{grumpy}) and (\ref{mcalS Cauchy problem}) it is manifest that the pure tension solutions (\ref{backgroundthemlineapp}) in the new coordinate system are exactly the pure tension solutions we consider in our analysis (see section \ref{Pure tension solutions}). This motivates our assertion that the analysis of \cite{deRham:2010rw} and ours study exactly the same system, although in different coordinate systems.

\subsubsection{Perturbations}

In the perturbative study of \cite{deRham:2010rw} a 4D scalar-vector-tensor decomposition is considered, and the field which become a ghost for small background tensions is the trace part of the $\m\n$ components of the bulk metric perturbation, which for clarity we indicate with $\bar{\pi}(\bar{z}, \bar{y}, \bar{x}^{\cdot})$ (in \cite{deRham:2010rw} it is indicated with $\pi$). More precisely, the evaluation on the cod-2 brane (which we indicate here as $\bar{\pi}^{(4)}(\bar{x}^{\cdot}) = \bar{\pi}(0,0,\bar{x}^{\cdot})$) of the $\bar{\pi}$ field is a ghost in an effective 4D description when $\bla < \blac$, while it is healthy when $\bla > \blac$. Since the coordinate transformations $(\star)$ and $(\star \star)$ leave untouched the 4D coordinates, the $\pi$ field of our analysis and the $\bar{\pi}$ field of \cite{deRham:2010rw} are linked by the transformation
\beq
\label{vedi ti}
\bar{\pi} \big( \bar{X}^{\cdot} (X^{\cdot})\big) = \pi \big( X^{\cdot} \big)
\eeq
where we indicated collectively $\bar{X}^{\cdot} = (\bar{z}, \bar{y}, \bar{x}^{\cdot})$ and $X^{\cdot} = (z, y, \xd)$. As a consistency check, we can use the coordinate transformation of section \ref{The change of coordinates} to derive the bulk equation for $\bpi$, starting from the bulk equation $\big( \de_{z}^{2} + \de_{y}^{2} + \de_{\mu} \de^{\m} \big) \pi (\Xd) = 0$ in the bulk-based description. A tedious calculation leads to the equation
\beq
\label{tedious}
\bar{\Box}_4 \, \bpi + \frac{1}{\bNq} \, \bigg( \debzq \bpi - 2 \, \bNy \, \debz \deby \bpi + (1 + \bq) \debyq \bpi - \frac{(1 + \bq) \bNyp}{\bNq} \, \debz \bpi + \frac{(1 + \bq) \bNy \bNyp}{\bNq} \, \deby \bpi \bigg) = 0
\eeq
where we used the definitions $\bN (\bar{y}) \equiv \sqrt{1 + \bq - \bq \bep^2 (\bar{y})}$ and $\bNy (\bar{y}) \equiv \b \, \bep (\bar{y})$. The equation (\ref{tedious}) is exactly the bulk equation which one obtains by varying the bulk action for the $\bpi$ field in \cite{deRham:2010rw}
\beq
S_6 = \int \! d^6 \bar{X} \,\, \frac{3 \Msf}{2} \, \Big( \bN \, \bpi \, \bar{\Box}_5 \, \bpi - \frac{1}{\bN} \, \big( \mcalL_{\bar{n}} \bpi \big)^2 \Big)
\eeq
where $\mcalL_{\bar{n}} \equiv \debz - \bNy \deby$ .

In particular, the relation (\ref{vedi ti}) implies that the evaluation of the $\bar{\pi}$ field on the cod-2 brane $\bar{\pi}^{(4)}$ in the analysis of \cite{deRham:2010rw} is \emph{exactly} equal to the $\pi$ field evaluated on the cod-2 brane $\pi^{(4)}(\chd)$ in our analysis
\beq
\bar{\pi}^{(4)} = \pi^{(4)}
\eeq
which implies that the 4D effective actions for $\bar{\pi}^{(4)}$ and $\pi^{(4)}$ have to be the same. Therefore, the values of the critical tension obtained using the two approaches have to agree.

\section{Thin limit of the background solutions}
\label{Thin limit of the background solutions}

In this appendix we clarify the properties of the thin limit of the background configurations, which we refer to in the main text. Since these solutions have been studied in detail in \cite{Sbisa':2014uza}, we refer to that paper for a more detailed discussion while we concentrate here only on the aspects which are more relevant for the present paper.

\subsection{The sequence of background solutions}

As we mention in the main text, the thin limit of the background is performed by considering a sequence of source configurations of the form
\beq
\label{angelicabackground}
\bar{T}^{[n]}_{ab}(\hxi, \chd) = - \d_{a}^{\, \, \m} \, \d_{b}^{\, \, \n} \, f_{1}^{[n]} \big( \hxi \big) \, \bla \,\, \g^{[n]}_{\m\n}(\hxi;\chd) - f^{[n]}_2(\hxi) \,\, \d_{a}^{\, \, \m} \,\, \d_{b}^{\, \, \n} \,\, \Mfs \, \mcal{G}^{[n]}_{\m\n}(\hxi;\chd)
\eeq
where $f_{1}^{_{[n]}}$ and $f_{2}^{_{[n]}}$ are sequences of even functions which satisfy
\begin{align}
\int_{-\infty}^{+\infty} f_{12}^{[n]} \big( \hxi \big) \, d \hxi &= 1 & f_{12}^{[n]} \big( \hxi \big) &= 0 \,\,\,\, \text{for} \,\,\,\, \abs{\hxi} \geq l^{[n]}_2
\end{align}
and $l^{_{[n]}}_2$ is a sequence of positive numbers that converges to zero: $l^{_{[n]}}_2 \rightarrow 0^+$ for $n \rightarrow + \infty$. There exist exact solutions for this class of sources \cite{Sbisa':2014uza} such that the bulk, induced and double induced metrics are respectively the 6D, 5D and 4D Minkowski metric, while the embedding of the cod-1 brane is $n$-dependent and non-trivial
\beq
\label{Pattysequence}
\bar{\varphi}_{[n]}^{A}(\hxi, \chd) = \big( Z_{[n]}(\hxi), Y_{[n]}(\hxi), \ch^{\a} \big) 
\eeq
and the cod-2 brane sits at $\hxi = 0$
\beq
\label{Robertsequence}
\bar{\a}_{[n]}^{a}(\chi^{\cdot}) = \big( 0, \chi^{\a} \big) 
\eeq
Imposing the condition $Z_{[n]}(0) = Y_{[n]}(0) = 0$, which implies that $Z_{[n]}(\hxi)$ is even while $Y_{[n]}(\hxi)$ is odd, we get \cite{Sbisa':2014uza}
\begin{align}
Z_{[n]}(\hxi) &= \int_{0}^{\hxi} \sin \bigg( \frac{\bla}{2 \Msf} \, \ep_{[n]}(\z) \bigg) d \z \label{Balotellibis1} \\[2mm]
Y_{[n]}(\hxi) &= \int_{0}^{\hxi} \cos \bigg( \frac{\bla}{2 \Msf} \, \ep_{[n]}(\z) \bigg) d \z \label{Balotellibis2}
\end{align}
where the regulating function is
\beq
\ep_{[n]}(\hxi) \equiv \int_{0}^{\hxi} f_{1}^{[n]} (\z) \, d \z 
\eeq
and in particular it vanishes in $\hxi = 0$ and is equal to $\pm \frac{1}{2}$ for $\hxi \gtrless \pm l_{2}^{_{[n]}}$.

\subsection{Thin limit of the embedding functions}

It is possible to see that, for $\hxi \gtrless l_2^{_{[n]}}$, we can write $Z_{[n]}(\hxi)$ and $Y_{[n]}(\hxi)$ as
\begin{align}
Z_{[n]}(\hxi) &= \sin \bigg( \frac{\bla}{4 \Msf} \bigg) \, \abs{\hxi} + Z_{[n]}^0 & Y_{[n]}(\hxi) &= \cos \bigg( \frac{\bla}{4 \Msf} \bigg) \, \hxi \pm Y_{[n]}^0
\end{align}
where \cite{Sbisa':2014uza}
\beq
\lim_{n \to \infty} Z_{[n]}^0 = \lim_{n \to \infty} Y_{[n]}^0 = 0
\eeq
Taking the limit $n \to \infty$ we get the thin limit configurations for the embedding functions
\begin{align}
Z(\hxi) &= \sin \bigg( \frac{\bla}{4 \Msf} \bigg) \, \abs{\hxi} & Y(\hxi) &= \cos \bigg( \frac{\bla}{4 \Msf} \bigg) \, \hxi
\end{align}
where now $\hxi$ is defined on all the real axis.

Concerning the first derivative with respect to $\hxi$ of the embedding functions, for every fixed value $\hxi$ different from zero (say positive, although the case $\hxi < 0$ is analogous), there exists a natural number $N$ such that, for $n \geq N$, we have $l^{_{[n]}}_2 < \hxi$ (as a consequence of $l^{_{[n]}}_2 \rightarrow 0$) and so $\ep_{^{[n]}}(\hxi) = 1/2$. Therefore (\ref{Balotellibis1}) implies 
\beq
\Zp_{[n]}(\hxi > 0) \xrightarrow[n \rightarrow + \infty]{} \sin \bigg( \frac{\bla}{4 \Msf} \bigg)
\eeq
On the other hand, again using (\ref{Balotellibis1}) we conclude that $\Zp_{[n]}(0) = 0$ independently of $n$, and so $\Zp_{^{[n]}}$ converges to
\beq
\label{felicita1}
\Zp \big( \hxi \big) =
\begin{cases}
\sin \Big( \bla/4 \Msf \Big) & \text{for $\hxi > 0$} \\
0 & \text{for $\hxi = 0$} \\
- \sin \Big( \bla/4 \Msf \Big) & \text{for $\hxi < 0$}
\end{cases}
\eeq
Analogously, (\ref{Balotellibis2}) implies that $\Yp_{[n]}(0) = 1$ independently of $n$, while for every fixed $\hxi \neq 0$ we have
\beq
\Yp_{[n]}(\hxi \neq 0) \xrightarrow[n \rightarrow + \infty]{} \cos \bigg( \frac{\bla}{4 \Msf} \bigg) 
\eeq
and so $\Yp_{^{[n]}}$ converges to
\beq
\label{felicita2}
\Yp \big( \hxi \big) =
\begin{cases}
\cos \Big( \bla/4 \Msf \Big) & \text{for $\hxi \neq 0$} \\
1 & \text{for $\hxi = 0$} 
\end{cases}
\eeq

\begin{figure}[htp!]
\begin{center}
\includegraphics{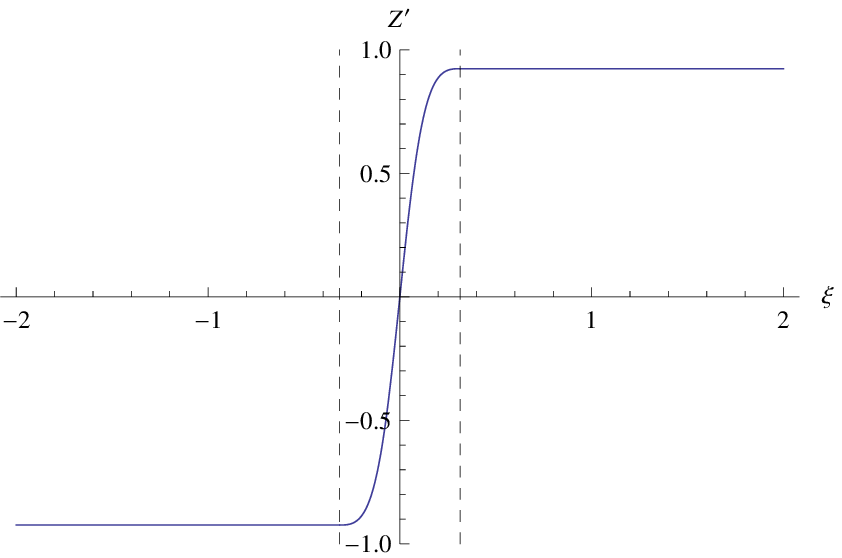}
\caption[The background embedding function $\Zp$]{Numerical plot of the background embedding function $\Zp_{[n]}$. The vertical dashed lines indicate the boundaries of the physical cod-2 brane.}
\label{background embedding Z}
\end{center}
\end{figure}

\begin{figure}[htp!]
\begin{center}
\includegraphics{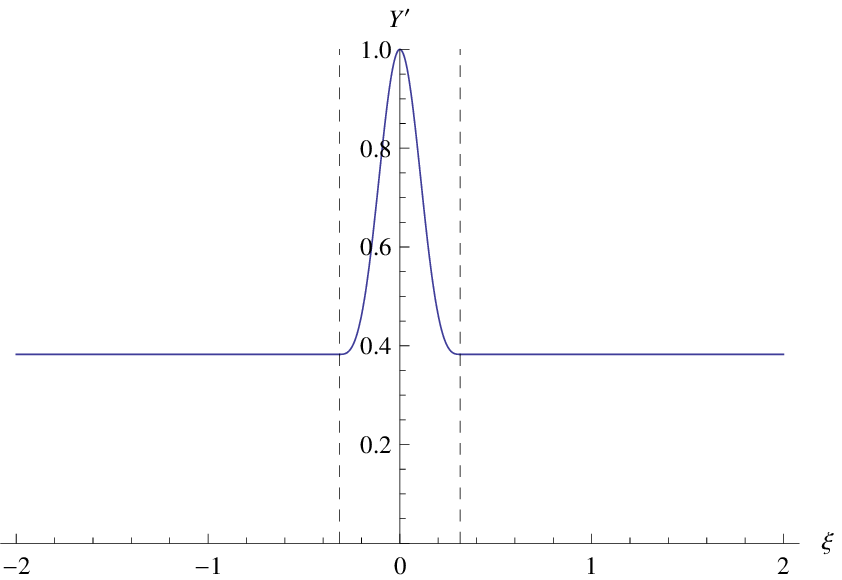}
\caption[The background embedding function $\Yp$]{Numerical plot of the background embedding function $\Yp_{[n]}$. The vertical dashed lines indicate the boundaries of the physical cod-2 brane.}
\label{background embedding Y}
\end{center}
\end{figure}

Note that both $\Zp$ and $\Yp$ are discontinuous. This is confirmed by the numerical plots figure \ref{background embedding Z} and \ref{background embedding Y} obtained in the pure tension perturbation case for $n = 10$ (using the explicit realizations (\ref{fnexplicit}) and (\ref{epnexplicit}), and same values for the free parameters as in section \ref{Choice of the free parameters}). As we mention in the main text, this has important consequences for the convergence properties of the sequences $\Zp_{^{[n]}}$ and $\Yp_{^{[n]}}$, since it implies that they cannot converge uniformly to their limit configurations. In fact, since $\Zp_{^{[n]}}$ and $\Yp_{^{[n]}}$ are smooth for $n$ finite, if they converged uniformly to $\Zp$ and $\Yp$ then the limit functions would necessarily be (at least) continuous.
In particular, the behaviour of $\Yp_{^{[n]}}$ is peculiar, since in the thin limit it has a removable discontinuity. This may suggest that, when we perform the pillbox integration, we can indeed remove it even for $n$ finite, justifying the use of the route B. However, from the point of view of the character of convergence of the sequence, the removable discontinuity has a crucial meaning: it signals that for every value of $n$ there is a peak of finite height at $\hxi = 0$. It is precisely this peak which is responsible for the mismatch between the results of the pillbox integration using the route A and the route B, since it gives a finite contribution to the pillbox integral for every value of $n$.

\section{Thin limit of a pure tension perturbation}
\label{Pure tension perturbations}

To study the thin limit of a pure tension perturbation, we introduce a sequence of perturbations of the cod-1 generalized energy-momentum tensor (\ref{deltaTmunu}) of the form
\beq
\label{marinapuretension}
\d\h{T}^{[n]}_{ab}(\hxi,\chd) = - f^{[n]}_1(\hxi) \, \dla \,\, \d_{a}^{\, \, \m} \,\, \d_{b}^{\, \, \n} \,\, \e_{\m\n} - f^{[n]}_2(\hxi) \,\, \d_{a}^{\, \, \m} \,\, \d_{b}^{\, \, \n} \,\, \Mfs \, \mcal{G}_{\m\n}(\hxi;\chd)
\eeq
where $\dla/\bla \ll 1$. The solution for the metric perturbations and bending modes in this case was obtained in \cite{Sbisa':2014uza} by solving the perturbative equations. However, the results of section \ref{Pure tension solutions} and of appendix \ref{Thin limit of the background solutions} imply that the exact (non-perturbative) solution when the tension on the cod-2 brane is $\bla + \dla$ is of the form
\begin{align}
g^{[n]}_{AB} &= \e_{AB} & \h{g}^{[n]}_{ab} &= \e_{ab} & g^{(4) \, [n]}_{\m\n} &= \e_{\m\n}
\end{align}
\beq
\vf^A(\hxi) = \Big( \mZ_{[n]}(\hxi), \mY_{[n]}(\hxi), 0,0,0,0 \Big)
\eeq
where $\mscr{Z}_{[n]}^{\prime}$ and $\mscr{Y}_{[n]}^{\prime}$ are expressed in terms of the regulating function as
\begin{align}
\mscr{Z}_{[n]}^{\prime}(\hxi) &= \sin \bigg( \frac{\bla + \dla}{2 M_6^4} \, \ep_{[n]} (\hxi) \bigg) & \mscr{Y}_{[n]}^{\prime}(\hxi) &= \cos \bigg( \frac{\bla + \dla}{2 M_6^4} \, \ep_{[n]} (\hxi) \bigg) 
\end{align}
We can then derive the perturbative solution by expanding the embedding functions at first order in $\dla$. In fact, at first order in $\dla$ we have
\begin{align}
\mscr{Z}_{[n]}^{\prime}(\hxi) &\simeq \sin \bigg( \frac{\bla}{2 M_6^4} \, \ep_{[n]} (\hxi) \bigg) + \cos \bigg( \frac{\bla}{2 M_6^4} \, \ep_{[n]} (\hxi) \bigg) \, \frac{\dla}{2 M_6^4} \, \ep_{[n]} (\hxi) \\[2mm]
\mscr{Y}_{[n]}^{\prime}(\hxi) &\simeq \cos \bigg( \frac{\bla}{2 M_6^4} \, \ep_{[n]} (\hxi) \bigg) - \sin \bigg( \frac{\bla}{2 M_6^4} \, \ep_{[n]} (\hxi) \bigg)\, \frac{\dla}{2 M_6^4} \, \ep_{[n]} (\hxi)
\end{align}
while for the background embedding we have
\begin{align}
Z_{[n]}^{\prime}(\hxi) &= \sin \bigg( \frac{\bla}{2 M_6^4} \, \ep_{[n]} (\hxi) \bigg) \\[2mm]
Y_{[n]}^{\prime}(\hxi) &= \cos \bigg( \frac{\bla}{2 M_6^4} \, \ep_{[n]} (\hxi) \bigg)
\end{align}
and so the $\hxi$-derivative of the $z$ and $y$ bending modes $\hdvf_{[n]}^{z \, \p} = \mscr{Z}^{\p}_{[n]} - Z^{\p}_{[n]}$ and $\hdvf_{[n]}^{y \, \p} = \mscr{Y}^{\p}_{[n]} - Y^{\p}_{[n]}$ read at first order in $\dla$
\begin{align}
\hdvf_{[n]}^{z \, \p}(\hxi) &\simeq \frac{\dla}{2 M_6^4} \, \ep_{[n]} (\hxi) \, \cos \bigg( \frac{\bla}{2 M_6^4} \, \ep_{[n]} (\hxi) \bigg) \\[2mm]
\hdvf^{y \, \p}_{[n]}(\hxi) &\simeq - \frac{\dla}{2 M_6^4} \, \ep_{[n]} (\hxi) \, \sin \bigg( \frac{\bla}{2 M_6^4} \, \ep_{[n]} (\hxi) \bigg)
\end{align}
Integrating with respect to $\hxi$, we obtain the $z$ and $y$ components of the bending modes at first order in $\dla$
\begin{align}
\hdvf_{[n]}^{z}(\hxi) &\simeq \d\!\b^{(4)}_{[n]} + \frac{\dla}{2 M_6^4} \, \int_{0}^{\hxi} \! d \z \, \, \ep_{[n]} (\z) \, \cos \bigg( \frac{\bla}{2 M_6^4} \, \ep_{[n]} (\z) \bigg) \\[2mm]
\hdvf^{y}_{[n]}(\hxi) &\simeq - \frac{\dla}{2 M_6^4} \, \int_{0}^{\hxi} \! d \z \, \, \ep_{[n]} (\z) \, \sin \bigg( \frac{\bla}{2 M_6^4} \, \ep_{[n]} (\z) \bigg)
\end{align}
where we used the fact that $\hdvf_{^{[n]}}^{y}(0)$ vanishes as a consequence of the $\mathbb{Z}_2$ symmetry which holds inside the cod-1 brane. Finally, we can construct the normal component of the bending $\hdvfn^{[n]} = \Yp_{[n]} \, \hdvf_{[n]}^{z} - \Zp_{[n]} \, \hdvf_{[n]}^{y}$ to get
\begin{align}
\label{heyheyappendix}
\hdvfn^{[n]}(\hxi) &\simeq \cos \bigg( \frac{\bla}{2 M_6^4} \, \ep_{[n]} (\hxi) \bigg) \, \d\!\b^{(4)}_{[n]} + \frac{\dla}{2 M_6^4} \, \cos \bigg( \frac{\bla}{2 M_6^4} \, \ep_{[n]} (\hxi) \bigg) \, \int_{0}^{\hxi} \! d \z \, \, \ep_{[n]} (\z) \, \cos \bigg( \frac{\bla}{2 M_6^4} \, \ep_{[n]} (\z) \bigg) + \nn \\[2mm]
& + \frac{\dla}{2 M_6^4} \, \sin \bigg( \frac{\bla}{2 M_6^4} \, \ep_{[n]} (\hxi) \bigg) \, \int_{0}^{\hxi} \! d \z \, \, \ep_{[n]} (\z) \, \sin \bigg( \frac{\bla}{2 M_6^4} \, \ep_{[n]} (\z) \bigg)
\end{align}
and in particular its value on the side of the cod-2 brane reads
\begin{align}
\label{enoughisenoughappendix}
\hdvfn^{[n]}\Big\rvert_{l_{2}^{[n]}} &\simeq \cos \bigg( \frac{\bla}{4 M_6^4} \bigg) \, \d\!\b^{(4)}_{[n]} + \frac{\dla}{2 M_6^4} \, \cos \bigg( \frac{\bla}{4 M_6^4} \bigg) \, \int_{0}^{l_{2}^{[n]}} \! d \z \, \, \ep_{[n]} (\z) \, \cos \bigg( \frac{\bla}{2 M_6^4} \, \ep_{[n]} (\z) \bigg) + \nn \\[2mm]
& + \frac{\dla}{2 M_6^4} \, \sin \bigg( \frac{\bla}{4 M_6^4} \bigg) \, \int_{0}^{l_{2}^{[n]}} \! d \z \, \, \ep_{[n]} (\z) \, \sin \bigg( \frac{\bla}{2 M_6^4} \, \ep_{[n]} (\z) \bigg)
\end{align}
Note that, once we specify the internal structure $f_{1}^{_{[n]}}$, the solution for the bending modes (and trivially for the metric) is known explicitly both inside and outside the thick cod-2 brane.

\end{appendix}

\clearemptydoublepage

\end{document}